On Using Medium-Range Ensemble Forecasts for Storm Transposition of Synoptic-Scale Systems in Probable Maximum Precipitation Estimation

Mathieu Mure-Ravaud

J. Amorocho Hydraulics Laboratory, Department of Civil and Environmental Engineering, University of California, Davis, 1 Shields Ave, Davis, 95616, California, USA.

## ABSTRACT

Most methods for estimating probable maximum precipitation (PMP)— the greatest depth of precipitation that is physically possible over a given area and duration—rely on storm transposition (ST), the process of transporting a storm, either historically observed or simulated, from its original location to a target basin. Existing ST approaches—whether classical or physically based—involve assumptions and manipulations that can introduce inconsistencies, leaving the physical validity of the transposed storm uncertain. In this study, the internal variability leveraging (IVL) approach is used to transpose an atmospheric river cluster that affected the U.S. West Coast during 20–29 October 2021. Steering the storm toward the target basin and determining its transposition region are achieved by considering an ensemble of plausible storm evolutions and trajectories obtained from archived ECMWF medium-range forecasts. The Willamette River and Nass River watersheds—located approximately 6° N, 2° W and 16° N, 8° W, respectively, from the area most affected by the observed precipitation—were selected as target basins. For each basin, the IVL realization yielding the largest 24-h basin-average precipitation depth was identified, and the initial and boundary condition shifting method was subsequently applied to further enhance its impact, producing 24-h precipitation depths of 119 mm for the Willamette and 98 mm for the Nass.

## LIST OF ACRONYMS

AMS – American Meteorological Society
AR – atmospheric river

BA – basin-average
CDF – cumulative distribution functions
DD – dynamical downscaling
ECMWF – European Centre for Medium-Range Weather Forecasts
ENS – ECMWF Ensemble Prediction System
ERA5 – fifth-generation ECMWF reanalysis
GCM – global circulation model
IBC – initial and boundary condition
IBCS – initial and boundary condition shifting
IC – initial condition
IVL – internal variability leveraging
IVT – integrated vapor transport
MCS – mesoscale convective system
NASEM – National Academies of Sciences, Engineering, and Medicine
NWP – numerical weather prediction
PB – physically based
PD – precipitation depth
PW – precipitable water
PMP – probable maximum precipitation
ST – storm transposition
UTC – Coordinated Universal Time
WRF – Weather Research and Forecasting

## I. Introduction

Concepts, definitions, and estimation procedures for probable maximum precipitation (PMP) have evolved continuously since the 1930s; a comprehensive account of this evolution is provided in Appendix B of the recent National Academies of Sciences, Engineering, and Medicine (NASEM, 2024) report. Among the recent formulations is the definition of the World Meteorological Organization (WMO, 2009): “the greatest depth of precipitation for a given duration meteorologically possible for a design watershed or a given storm area at a particular location at a particular time of year, with no allowance made for long-term climatic trends”. NASEM (2024) recommended revising the definition of PMP to: “the depth of precipitation for a particular duration, location, and areal extent, such as a drainage basin, with an extremely low annual probability of being exceeded, for a specified climate period”. This revision removes the assumption that rainfall is absolutely bounded, encourages the use of procedures that explicitly account for statistical uncertainty in PMP estimates, and incorporates climate dependence so that changes in extreme precipitation under climate change can be reflected.

PMP estimates are used primarily in hydrologic design and risk assessment for critical infrastructure, including dams, nuclear power plants, and flood control systems. They are essential for ensuring that reservoirs and spillways are designed to withstand extreme rainfall events, thereby reducing the risk of catastrophic flooding and dam failures. Several methodologies have been developed over the decades for estimating PMP. Statistical approaches include the Hershfield method (Hershfield 1961, 1965; Koutsoyiannis 1999), which has been widely applied in PMP estimation, and extreme value analysis (EVA; Ben Alaya et al. 2018; Coles 2001), which is commonly used to assess the annual exceedance probabilities (AEPs) of extreme rainfall. The other traditional PMP estimation approach—referred to by the WMO (2009) as the "hydrometeorological approach"—adopts a more physically based perspective, relying on a comprehensive analysis of the hydrometeorological mechanisms that produce extreme precipitation. It is built upon the following core components (Hansen, 1987; NASEM, 2024):

1. **Establishment of a catalog of extreme storms** that serve as the observational foundation for PMP estimation.
2. **Moisture maximization** of observed precipitation amounts to reflect the potential rainfall under more extreme atmospheric moisture conditions.
3. **Storm transposition (ST)**, whereby the maximized storm rainfall is relocated to the basin of interest. This relocation is acceptable only if the basin lies within the "storm transposition region"—the geographic domain where the storm is deemed meteorologically plausible.
4. **Application of transposition correction factors**, as needed, to adjust for differences in moisture availability or to account for orographic influences (e.g., terrain-induced enhancement or depletion of precipitation).
5. **Envelopment**, in which PMP curves are constructed by combining the largest values from all moisture-maximized and transposed storms

Details on the hydrometeorological approach and its application to specific sites or regions are documented in the series of Hydrometeorological Reports (HMRs) and Technical Papers (TPs) produced by the U.S. Weather Bureau (USWB) in collaboration with the U.S. Army Corps of Engineers (USACE) and the U.S. Bureau of Reclamation (USBR) during the mid-to-late 20th century (e.g., Corrigan et al., 1999; Hansen et al., 1982, 1988, 1994; Ho and Riedel, 1980; Schreiner and Riedel, 1978). The hydrometeorological approach was also reviewed in three WMO manuals on PMP estimation (WMO, 1973, 1986, 2009), in recent

NUREG reports (DeNeale et al., 2021; England et al., 2020), and most recently in the NASEM (2024) report.

Traditional PMP estimation approaches face important limitations. Statistical methods are constrained by the scarcity of extreme precipitation records, leading to large uncertainties in model parameters and, consequently, in PMP estimates (Klemeš, 1993, 2000a,b; Moran, 1957). The hydrometeorological approach, developed at a time when computational resources were insufficient to resolve the complex structures and processes of extreme-precipitation storms, relies on strong assumptions and involves substantial subjective judgment (Chen & Bradley, 2006, 2007; DeNeale et al., 2021). Several of these assumptions have since been demonstrated to be generally invalid through numerical modeling studies (Abbs, 1999; Chen, 2005; Ohara et al., 2017; Yang & Smith, 2018; Zhao et al., 1997), such as the assumption that, in intense storms, total precipitation and available moisture are linearly related. For these reasons, NASEM (2024) emphasized revisiting the concept of PMP and its estimation methodology, and advocated transitioning to computer-based approaches.

By numerically solving the governing equations for the conservation of mass, momentum, and energy, numerical weather prediction (NWP) models can reconstruct the full evolution of a storm system while explicitly accounting for topographical influences. NWP model-based methods—often referred to as “physically-based” (PB) methods—have emerged in recent years for PMP estimation and remain an active area of research. While recent applications of PB methods include tropical cyclones (Ishikawa et al., 2013; Oku et al., 2014; Mure-Ravaud et al., 2019a,b ) and mesoscale convective systems (MCSs, Hiraga & Tahara, 2025; Hiraga et al., 2025c; Kavvas et al., 2023; Mure-Ravaud & Kavvas, 2026), the majority of studies to date have focused on atmospheric rivers (ARs).

An AR is a long, narrow, and transient corridor of strong horizontal water vapor transport, typically associated with a low-level jet stream ahead of the cold front of an extratropical cyclone (Gimeno et al., 2014; Ralph et al., 2004, 2018). ARs are the primary mechanism by which water vapor is transported through the atmosphere poleward of the tropics (Ralph et al., 2005, 2020; Sodemann & Stohl, 2013; Zhu & Newell, 1998), and therefore play a crucial role in both global and regional water budgets (Dettinger, 2013; Guan et al., 2010; Paltan et al., 2017). When forced upward by orographic barriers or ascent within the warm conveyor belt, ARs frequently produce heavy precipitation that can trigger flooding and cause major socioeconomic impacts (Lavers & Villarini, 2013; Ralph et al., 2006, 2011, 2013).

The most widely used PB method for estimating PMP from ARs is that of Ohara et al. (2011). In this framework, ST is implemented by spatially shifting the gridded fields used as initial and boundary conditions (IBCs) in the NWP model to redirect the AR’s trajectory, which we will hereafter refer to as the “initial and boundary condition shifting” (IBCS) method, while moisture maximization is performed by prescribing 100% relative humidity along the model domain boundaries. This methodology was later refined by Ishida et al. (2015a,b) and Toride et al. (2019), who introduced a moisture maximization criterion based on an integrated vapor transport[1] (IVT) threshold, thereby maximizing relative humidity only along the AR pathway in order to preserve realistic storm and precipitation structures. Toride et al. (2019) demonstrated that prescribing 100% relative humidity at the model boundaries does not necessarily yield maximum precipitation over a target watershed, and therefore argued that PMP estimation should be approached as a moisture perturbation problem rather than a moisture maximization exercise.

Since the work of Ohara et al. (2011), numerous PMP studies have employed either the IBCS method alone (Hiraga et al., 2025b; Ishida et al., 2018; Liang & Yong, 2022), the moisture-perturbation method alone (Ohara et al., 2017; Rastogi et al., 2017; Tarouilly et al., 2023), or a combination of the two (Chen et al., 2016; Hiraga & Meza, 2025; Hiraga et al., 2021, 2023; Tarouilly et al., 2024; Trinh et al., 2022), targeting watersheds both within the United States and abroad.

The present study focuses on ST and builds upon the work of Mure-Ravaud & Kavvas (2026, hereafter MK26). Section 2 describes how medium-range ensemble forecasts can be leveraged to support ST. Section 3 details the numerical setup for dynamical downscaling (DD) with the Weather Research and Forecasting (WRF) model[2] and presents the IBCS results. Section 4 summarizes the key findings and conclusions.

## II. Letting the atmosphere decide: internal variability as a driver of storm transposition

---

[1] The integrated vapor transport (IVT) field—also known as the moisture advection field—is a two-dimensional vector field that quantifies the horizontal transport of water vapor by the wind. It is defined as:

$$\mathrm{IVT} = \int_{z=0}^{z_{top}} \rho_v \, \mathbf{U} \, dz = \int_{z=0}^{z_{top}} q_v \, \rho \, \mathbf{U} \, dz$$

where z is the vertical coordinate (height), $z_{top}$ is the height of the top of the atmosphere, $\rho_v$ is the water vapor density, $q_v$ is the specific humidity, $\rho$ is the air density, $\mathbf{U}$ is the horizontal wind vector.

[2] The WRF model is a state-of-the-art NWP system designed for both atmospheric research and operational forecasting (Powers et al. 2017; Skamarock & Klemp 2008; Skamarock et al. 2019). This fully compressible, non-hydrostatic model provides a flexible and computationally efficient framework for simulating atmospheric processes across a wide range of spatial and temporal scales. WRF includes advanced physics parameterizations for cloud microphysics, radiation, boundary-layer dynamics, and land–surface interactions, making it suitable for applications spanning short-term weather prediction to long-term climate research.

IBCS is a powerful tool for PB ST: contrary to traditional ST approaches, where the observed rainfall pattern (isohyetal contours or gridded precipitation fields) is simply shifted geometrically—i.e., without dynamically recalculating the atmosphere—to the target watershed, the IBCS method instead transposes the storm by spatially shifting the IBC gridded fields, allowing the NWP model to freely evolve under its governing equations, and thus to produce a precipitation field that is dynamically consistent with the atmospheric circulation and responsive to the underlying topography.

However, this method also has limitations, discussed in MK26, which have received little attention so far. One key issue arises from the spatial non-stationarity of atmospheric fields, meaning that the range of values a given variable can take may vary markedly across space. This is exemplified in Fig. 1a, which shows the meridional wind speed from the fifth-generation European Centre for Medium-Range Weather Forecasts reanalysis (ERA5; Hersbach et al. 2023a,b) at 1000 hPa on 21 October 2021 00:00 UTC. As expected, wind speeds tend to be higher over the smoother surfaces of oceans and lakes than over the rougher land surface. Consequently, shifting this field (Fig. 1b) transports stronger overwater winds inland while moving weaker overland winds offshore. Such inconsistencies arise for nearly all variables and are particularly evident near the surface. Hence, shifting the IBCs may effectively bring "the sea over land" or "mountains over plains," introducing atmospheric conditions that are inconsistent with the local environment. Some studies (e.g., Toride et al., 2019) sought to address this issue by constraining the shifts to within ±1° longitude and ±5° latitude. The choice of these bounds, however, remains arbitrary and does not guarantee physical consistency. Additionally, the target area may lie beyond the prescribed bounds.

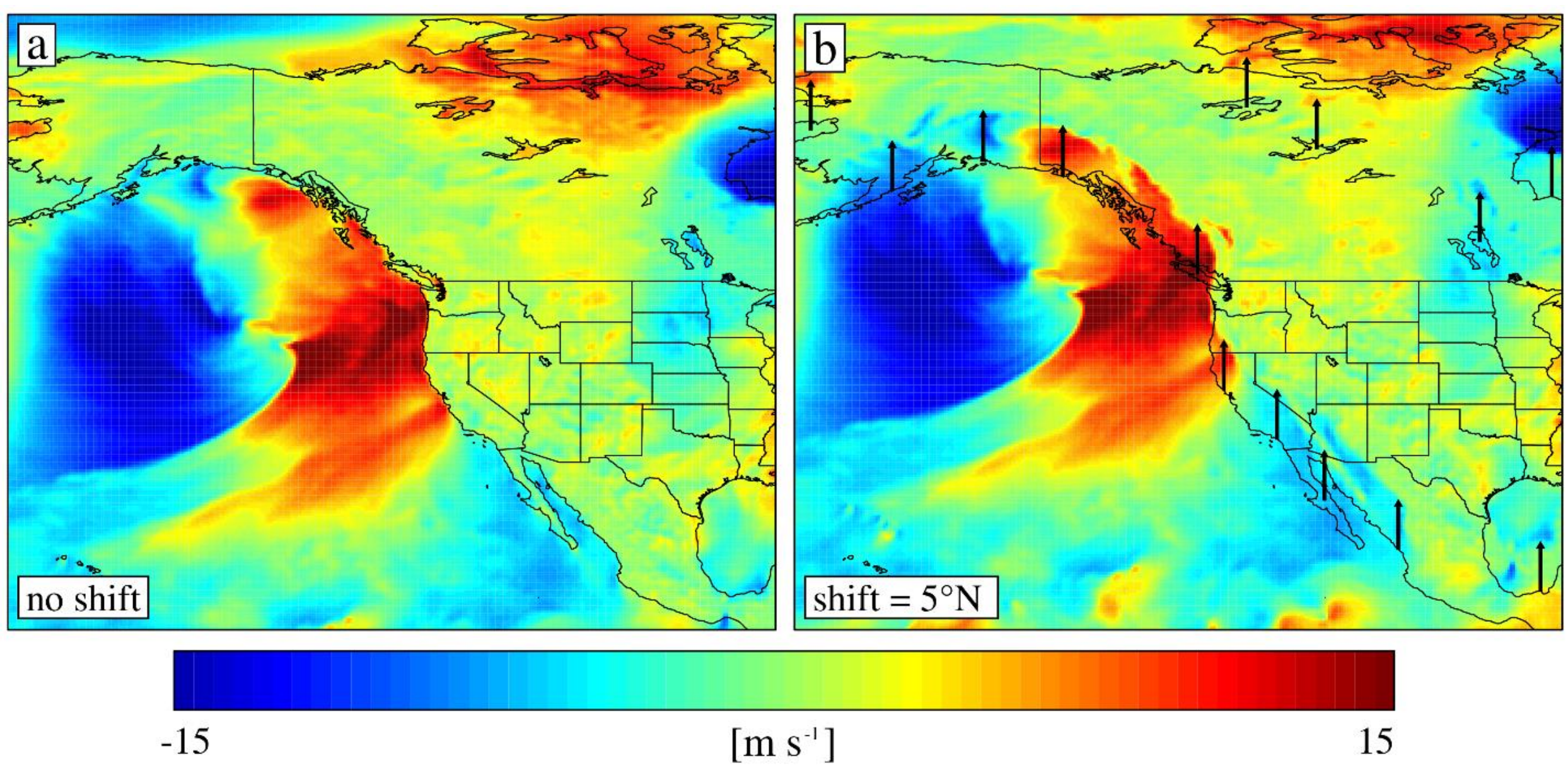

**Figure 1** – Meridional wind speed (m $s^{-1}$) from ERA5 at 1000 hPa on 21 Oct 2021 00:00 UTC for (a) no shift and (b) a 5°N shift. Black arrows are used in (b) to highlight inconsistencies introduced by the shifting.

When attempting to make a storm occur at a different location, the most immediate and computationally inexpensive expedient is to shift the PD field itself, as in the traditional ST approach. A more physically grounded advancement is the IBCS method, which manipulates the model's IBCs to steer the storm over the target basin. Despite this improvement, however, both methods rest on the same fundamental premise—that storm placement can be prescribed by spatially displacing the atmospheric fields. Yet, a more physically consistent alternative exists—one that can not only steer the storm toward the target basin but also help delineate the ST region, offering an objective alternative for defining this region compared to traditional approaches that rely heavily on subjective meteorological judgment.

The atmosphere, being a nonlinear dynamical system, exhibits complex behavior in which even small perturbations to the IBCs can amplify unpredictably and lead to vastly different outcomes over time (Lorenz 1963, 1969; Madden 1976; Thompson 1957). MK26 leveraged this sensitivity to transpose an MCS over the Midwestern U.S. and to construct its ST region—an approach they termed the "internal variability leveraging" (IVL) method. More specifically, an ensemble of MCS realizations was generated by incrementally advancing the simulation start time and expanding the outer domain boundaries, thereby allowing the WRF model to explore a wide range of plausible storm evolutions and trajectories. This approach is illustrated schematically in Fig. 2a. The union of the rainfall footprints from all ensemble realizations delineates the spatial domain within which the storm can physically occur—that is, the ST region. Because the spatial extent of this ensemble-derived ST region naturally expands with increasing ensemble size, fully capturing it requires generating a sufficiently large ensemble. Given the computational expense of producing new simulations, the ensemble generation strategy is therefore particularly important. Moreover, if IVL fails to yield any realization affecting the target basin, this indicates that the storm's displacement toward that location is not dynamically supported, and applying ST in such a case would lack physical justification.

In MK26, IVL represents only the first step of the ST process. It is unlikely that any single realization will produce the maximum impact over the target basin, as the simulated PD field may fall slightly outside the basin or only partially overlap with it. Hence, it becomes necessary to reintroduce IBCS—not as the primary transposition tool, as in previous studies, but as a means of fine-tuning the storm's position to maximize the basin-average (BA) PD. In this framework, the applied IBC shifts are relatively small compared to cases where IBCS served as the main transposition mechanism and can therefore be regarded as minor perturbations to the IBCs used to drive the regional simulations. Consequently, the potential influence of IBCS on the physical consistency of the simulated fields, as discussed earlier, is substantially reduced. Nonetheless, while IVL reduces the magnitude of the shift required by IBCS, MK26—like previous studies—did not provide guidance on what constitutes an acceptable maximum IBC shift. In operational contexts, such decisions will likely remain at the discretion of the hydrometeorologist or design engineer and will need to be made on a case-by-case basis.

In addition to ensemble size and construction strategy, another key consideration is to ensure that the model's boundaries are not overly restrictive. The model must be afforded sufficient spatial freedom for internal variability to fully express itself and for the storm to explore the full range of physically plausible trajectories. In MK26, the MCS's relatively limited spatial footprint within the model domains, combined with its high sensitivity to the simulation start time and outer boundary placement, meant that these boundaries imposed minimal constraint on the storm's ability to explore different regions of the inner domain. For more predictable, synoptic-scale systems such as ARs, however, the situation differs: the limited spatial extent of a regional model inherently constrains the large-scale dynamics, leaving little opportunity for substantial spread in storm evolution scenarios. This contrast is exemplified by Tarouilly et al. (2023), who generated ensembles of WRF simulations by varying combinations of physical parameterizations and activating the model's stochastic parameterization schemes to assess the sensitivity of 72-h PMP estimates over a California watershed to physical parameterizations, IBC error, and upscale-propagating model uncertainties. They found a relatively narrow spread in the maximized 72-h BA PD across ensemble members—approximately ±7% of the ensemble mean.

Achieving a larger ensemble spread requires introducing perturbations further upstream in the modeling chain, as demonstrated by Ødemark et al. (2021). They generated an ensemble of dynamically downscaled simulations of an AR event and its associated extreme precipitation over two watersheds in western Norway by altering the IBCs. In this case, Ødemark et al. (2021) leveraged the internal variability not of the regional model, but of the global circulation model (GCM) that provided those IBCs. Unlike MK26, this approach did not rely on changing the simulation start time; instead, stochastic perturbations were added to the GCM's physics tendencies to represent uncertainties in the parameterization of subgrid-scale physical processes (Buizza et al. 1999).

Ødemark et al. (2021) noted that their objective was "to investigate whether the ensemble approach can serve as an alternative to the physically more inconsistent artificial manipulations of initial and boundary conditions [i.e., IBCS and moisture perturbation] in the NWP model." However, as the present study demonstrates, these approaches are in fact highly complementary for PB PMP estimation and need not be regarded as mutually exclusive. The IVL-based ensemble performs the "heavy lifting" of steering the storm toward dynamically plausible new locations and delineating the ST region, whereas IBCS serves to maximize the storm's impact over a specific target basin by fine-tuning its position.

In their conclusion, Ødemark et al. (2021) proposed an idea that underpins the present study: "To meet these challenges [i.e., the number of extreme storm simulations associated with ensemble-based PMP estimation] with realistic computational costs, a possible approach could be to utilize already established ensembles from numerical seasonal or weather forecasting systems".

Wick et al. (2013) evaluated the ability of five operational ensemble forecast systems to represent and predict ARs as a function of lead time out to 10 days over the northeastern Pacific Ocean and the west coast of North America. They found that average landfall position errors reached over 800 km at a 10-day lead time. These results suggest that transposing an AR to a target watershed located several hundred kilometers away would require letting the GCM evolve for at least a few days for internal variability to play out.

Forecasting systems are generally classified according to their prediction lead time, ranging from nowcasting (up to a few hours) and short-range (up to 2–3 days) to medium-range (up to ~2 weeks), subseasonal (up to ~6 weeks), and seasonal (up to ~6–12 months) forecasts (Buizza, 2019; Sun et al., 2014; Vitart and Robertson, 2019). The forecast horizons relevant to this study therefore correspond primarily to the medium-range, and possibly to the subseasonal range up to 2–3 weeks. Major global medium-range ensemble prediction systems include, among others, the European Centre for Medium-Range Weather Forecasts (ECMWF) Ensemble Prediction System (ENS; Buizza et al. 2007, 2008; Molteni et al. 1996), the National Centers for Environmental Prediction (NCEP) Global Ensemble Forecast System (GEFS; Toth and Kalnay 1997; Zhou et al. 2017), the UK Met Office Global and Regional Ensemble Prediction System (MOGREPS; Bowler et al. 2008; Inverarity et al. 2023), and the Canadian Global Ensemble Prediction System (GEPS; Charron et al. 2010; Lin et al. 2016).

Building upon the idea proposed by Ødemark et al. (2021), ST is implemented in this study using realizations drawn from the ECMWF ENS—an ensemble-based probabilistic forecasting framework designed to represent the range of possible weather conditions up to 15 days ahead. ENS comprises 51 forecasts: one control member (CNTL) and 50 perturbed members that differ slightly in their initial conditions and model physics.

The initial intent of this study was to perform the ST of an extreme AR that impacted the West Coast of North America on 24–25 October 2021. This storm was the strongest October AR to make landfall in the San Francisco Bay Area in the past four decades and produced AR3–AR5 conditions (based on the Ralph et al. (2019) AR intensity scale[3]) across coastal Central and Northern California, with IVT values locally exceeding 1,200 kg $m^{-1}$ $s^{-1}$ (CW3E, 2021). This AR was associated with one of the most intense extratropical cyclones on record to affect the Pacific Northwest—often referred to as a “bomb cyclone” (Sanders & Gyakum, 1980; Zhu & Newell, 1994)—which underwent a pressure drop of over 40 hPa in 24 hours. The event was well forecast up to 3 days in advance, with the presence of the AR accurately predicted out to 6 days. Moreover, subseasonal forecasts indicated enhanced AR activity as early as 3 weeks prior to the event (CW3E, 2021). This is consistent with our earlier discussion on the relevance of medium-range forecasts for constructing the IVL ensemble. The 24–25 October AR was in fact the fourth and most intense in a sequence of six ARs that made landfall along the U.S. West Coast between 20 and 29 October. This AR cluster brought high winds and extreme precipitation that, on the one hand, helped extinguish wildfires and provided some relief from the severe drought affecting Central and

[3] The Ralph et al. (2019) AR intensity scale aims to characterize atmospheric river strength and impacts in a way that is both useful to scientists and accessible to nonexperts. It is based on the maximum instantaneous IVT magnitude and the duration of AR conditions—two key characteristics that largely determine streamflow magnitudes and related hydrologic impacts. The scale is analogous in concept to hurricane categories, ranging from AR-1 (weakest) to AR-5 (strongest).

Northern California, but on the other hand, caused flooding, power outages, and landslides (NASA Earth Observatory, 2021). Some locations in Northern California received PDs exceeding 15 in from this sequence of storms. The 24–25 October AR alone produced more than 10 in of precipitation in parts of the Northern California Coast Ranges and the northern Sierra Nevada, and over 3 in across portions of the higher terrain in Nevada and Idaho (CW3E, 2021).

Our original plan was to transpose the single AR event of 24–25 October 2021 using the IVL–IBCS approach, leveraging the fact that increasing forecast lead times naturally allow internal variability to manifest as progressively distinct AR trajectories and landfall locations. As the lead time increases, however, this variability begins to modify not only the landfall position but also the timing—and eventually even the number, internal structure, and relative intensities (i.e., the 24–25 October AR may no longer remain the strongest)—of the ARs within the cluster. In other words, at short leads, a one-to-one correspondence can still be drawn between the observed and forecasted ARs, whereas beyond a certain point the large-scale “ingredients”—that is, the large-scale dynamical and thermodynamical precursors (e.g., the positioning and intensity of the parent extratropical cyclone and the jet stream configuration)—reorganize into a different yet dynamically consistent AR cluster. The model therefore no longer reproduces the 24–25 October event and its parent cluster exactly, but instead generates a physically plausible parallel sequence. Consequently, isolating and transposing only the 24–25 October AR is no longer physically meaningful; rather, the entire AR cluster must be treated as the operative unit for transposition. This implies that the transposition must encompass the full dynamical envelope of the AR cluster, allowing the model to explore a wide range of internally consistent realizations in which the large-scale ingredients reorganize in different ways—some of which may ultimately produce the most extreme impact over the target basin.

In this framework, though, can one still meaningfully speak of “storm” transposition? This raises the broader question of what a storm actually is. According to the *Cambridge Dictionary*, a storm is defined as “an extreme weather condition with very strong wind, heavy rain, and often thunder and lightning”[4]. In this sense, the term refers to the final manifestation—downstream in the causal chain—of the unfolding of the large-scale atmospheric ingredients described above. Under that definition, the process employed here would not strictly qualify as “storm transposition”, whereas traditional approaches that directly shift the observed precipitation would.

However, the American Meteorological Society (AMS) Glossary of Meteorology defines a storm more broadly as “a disturbed state of Earth's atmosphere which can manifest itself

[4] https://dictionary.cambridge.org/us/dictionary/english/storm (Accessed March 30, 2026)

in temperature, humidity, pressure, wind velocity, cloud cover, lightning, and precipitation", and as an "organized disturbance"[5], with "disturbance" itself defined as "any agitation or disruption of a steady state"[6]. This broader definition implies that the planetary- and synoptic-scale "ingredients" (i.e., disturbances) that ultimately give rise to extreme precipitation may themselves be regarded as the storm. In other words, within the framework of the AMS definition, the generation of realizations representing alternative unfoldings of these large-scale ingredients —among which at least one produces a dynamically consistent impact over the target basin—can legitimately be interpreted as "storm transposition" as well.

Medium-range forecasts for the 20–29 October AR cluster were retrieved from the operational archive of ECMWF's Meteorological Archival and Retrieval System (MARS). The downloaded variables included precipitation, precipitable water (PW), and integrated vapor transport (IVT). For the realizations selected for DD (see Section 3), we additionally retrieved the three-dimensional atmospheric fields required by WRF—air temperature, relative humidity, geopotential height, and wind components—available on 12 pressure levels from 1000 hPa down to 10 hPa. The near-surface fields (surface pressure, mean sea-level pressure, 10-m wind speed, 2-m air temperature, and 2-m dew-point temperature) and land-surface fields (soil moisture, soil temperature, skin temperature, surface geopotential, and land–sea mask) were also extracted. The data are provided on the octahedral reduced Gaussian grid O640 (~18 km), which corresponded to the horizontal resolution of the Integrated Forecasting System (IFS)—ECMWF's global NWP model—in October 2021. The medium-range ensemble forecasts are produced four times daily at 00, 06, 12, and 18 UTC. The 00 and 12 UTC cycles extend to 15 days, whereas the 06 and 18 UTC cycles extend to 6 days.

To ensure full coverage of the AR cluster while accommodating possible timing discrepancies arising from internal variability among realizations, precipitation forecasts for the period 19–30 October were used in this study. Given the previously discussed predictability horizon of ARs, forecasts initialized after 22 October were excluded, as the simulated event would be too similar to the observed one and therefore not relevant to the ST objectives of this study. In addition, all forecasts were required to include at least the interval 19–22 October. This criterion excludes realizations with early initialization times that have little temporal overlap with the analysis period. This selection resulted in 43 forecast initialization times, spanning 7 October 00 UTC to 22 October 00 UTC, and corresponding to 43 × 51 = 2193 realizations of the AR cluster (50 perturbed members plus one control member per initialization time). The approach proposed herein thus enables the construction of a large IVL ensemble directly from archived forecast data, without the need for additional simulations. It is illustrated schematically in Fig. 2b. In contrast, in MK26, generating the 135-member IVL ensemble for the MCS case required weeks of computation on several high-performance computers.

[5] https://glossary.ametsoc.org/wiki/storm (Accessed March 30, 2026)
[6] https://glossary.ametsoc.org/wiki/disturbance (Accessed March 30, 2026)

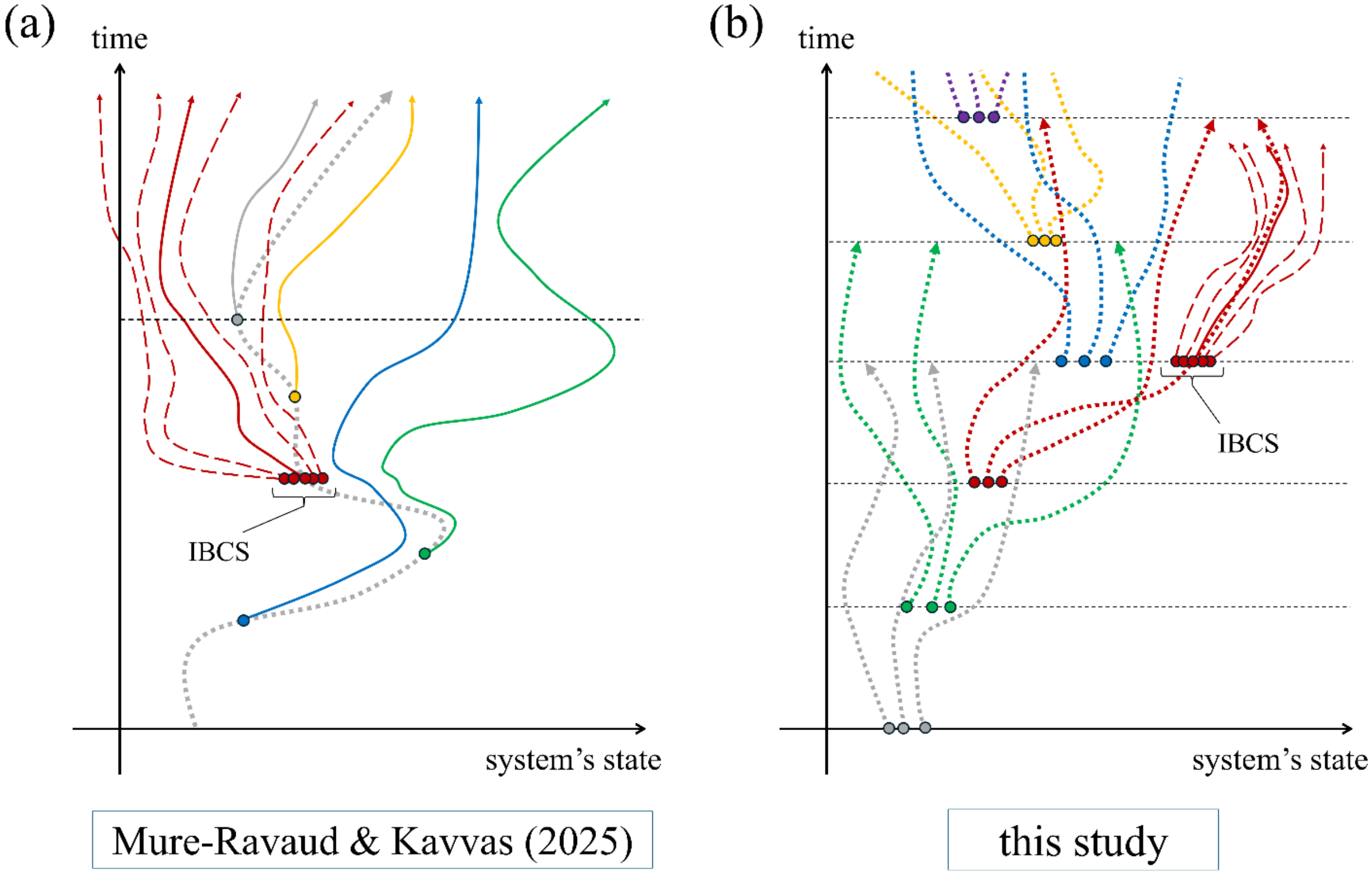

**Figure 2.** Schematic representation of ST procedures in (a) Mure-Ravaud & Kavvas (2026) and (b) the present study. Thick dotted lines denote the large-scale forcing fields used as IBCs for the downscaled simulations: ERA5 in (a) and ECMWF medium-range ensemble forecasts in (b). Dashed lines indicate IBCS runs—that is, WRF simulations corresponding to shifted IBCs—while solid lines represent WRF simulations with no IBC shift. Note that, in the schematic, IBCS is applied to a single IVL realization, consistent with MK26 and with the present study, in which it was applied only to the realization that produced the largest BA PD. In (a), the thin black horizontal dashed line marks the control-run start time, whereas in (b), thin black horizontal dashed lines indicate the 6-hourly forecast cycles, i.e., the times when new forecasts are initialized from updated analyses.

Composite PD fields were constructed by calculating, at each grid point, the maximum PD across all realizations, for various accumulation durations. These fields are shown in Fig. 3a–c. They effectively represent the spatial footprint of the AR, or ST region. Here, "ST region" is to be understood in the broader sense introduced earlier: not as the area to which a rainfall field can be artificially displaced, but as the region that may ultimately experience precipitation resulting from the unfolding of the storm's large-scale atmospheric ingredients.

Whereas in the MCS case analyzed in MK26 these composite PD fields exhibited a fairly well-defined outer envelope—i.e., PD values decreased to near zero when moving away from the region—here, the AR cluster appears capable of influencing a much larger portion of North America, and no clear external limit of its region of influence can be identified within the geographical domain shown in the figure. In fact, as one moves farther eastward, PD values eventually increase again (not shown in Fig. 3) because, during the same period, the U.S. East Coast was impacted by a powerful nor'easter—which later evolved into Tropical Storm Wanda (Reinhart & Berg, 2022)—that also brought substantial precipitation. As expected, Fig. 3a–c additionally reveals a pronounced PD gradient from coastal to inland regions, reflecting the orographic enhancement of precipitation during AR landfall.

It is important to note that while these composite fields provide a meaningful depiction of the AR cluster's spatial footprint, their quantitative reliability remains contingent on the accuracy of the underlying ECMWF precipitation forecasts. To assess this accuracy, Fig. 3d–f shows the same pointwise maximum PDs, but constructed using only precipitation from the control forecast (i.e., the unperturbed member) at 0–6 h lead times. The 0–6 h interval is used because PD is an accumulated quantity and the 0-h field is therefore always zero. Comparison of Fig. 3d–f with Fig. 3g–i, which displays the pointwise maxima derived from Stage IV[7] observations, indicates that the ECMWF IFS tends to underestimate precipitation from this AR cluster, although the location and overall pattern of the composite PD field are well captured. Consequently, the intensity of the composite PD fields constructed using the entire ensemble (Fig. 3a–c) is likely underestimated, whereas their spatial structure (e.g., the pronounced PD gradient from coast to inland regions) can reasonably be considered reliable. Furthermore, Fig. 3a–c reveals that the IVL ensemble includes realizations producing extreme precipitation in regions far from those impacted by the observed AR cluster, thereby illustrating the potential of the proposed approach to achieve fully physically consistent ST even for target basins located well beyond the ±1° longitude and ±5° latitude IBC shift limits used by Toride et al. (2019) and subsequent IBCS-based PMP studies.

[7] Stage IV is a mosaic of regional multi-sensor precipitation analyses produced by the National Weather Service (NWS) River Forecast Centers (RFCs) since December 2001. It combines rain gauge data with radar-estimated rainfall and is available at a 4-km horizontal resolution with three temporal resolutions: 1 h, 6 h, and 24 h (Du, 2011; Lin & Mitchell, 2005; Nelson et al., 2016).

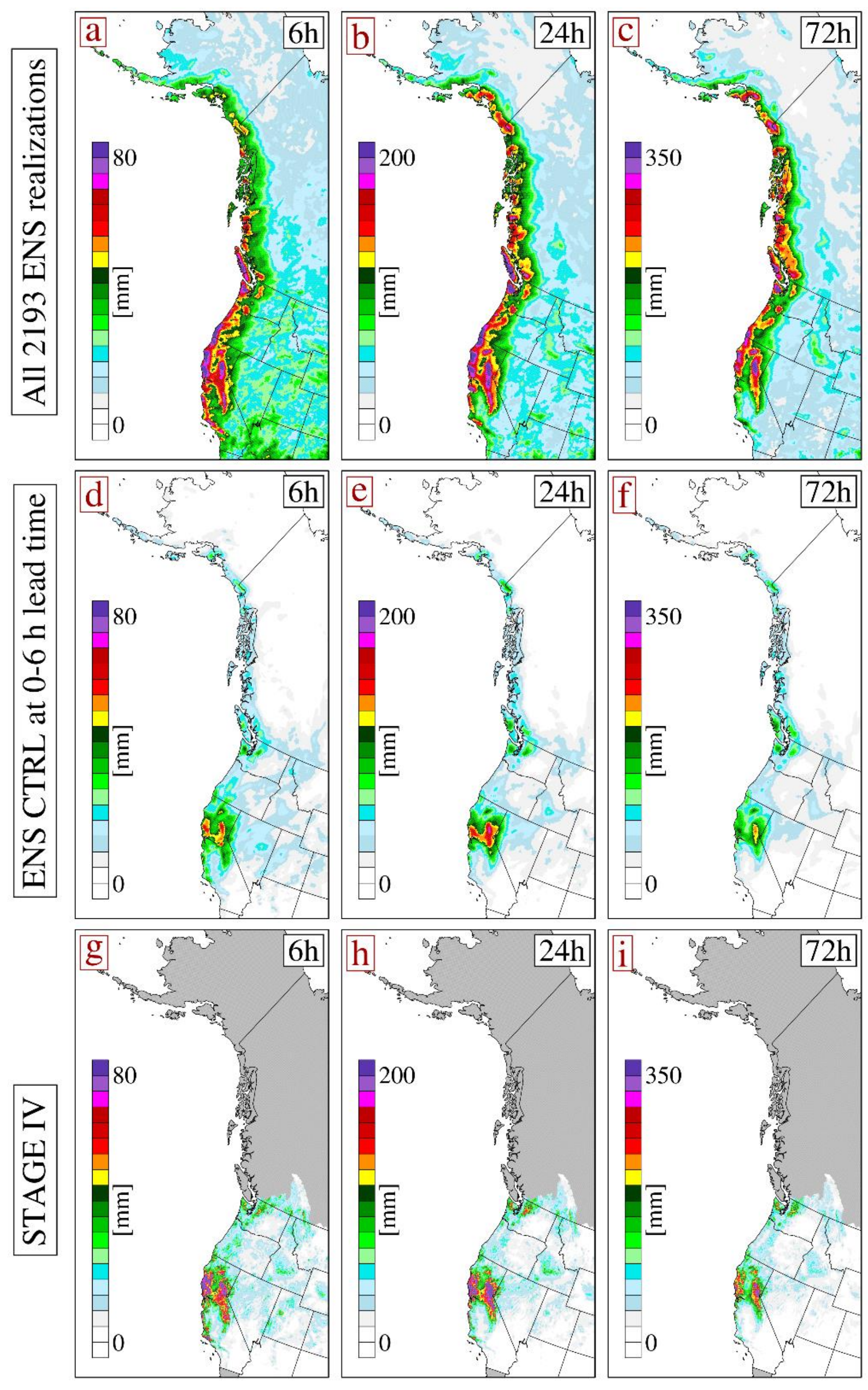

**Figure 3** – Pointwise maximum N-hour accumulated PD (mm), where N = 6 (left column), 24 (middle column), and 72 (right column). The top row shows the maxima across all 2193

realizations of the AR cluster from the ECMWF medium-range ensemble forecasts. The middle row shows the pointwise maximum N-hour PDs computed from the control forecast at 0–6 h lead time*, used to assess the accuracy of ECMWF precipitation estimates against Stage IV observations. The bottom row shows the same pointwise maxima calculated from Stage IV observations. At each grid point, the maximum N-hour PD was first calculated by sliding a continuous (i.e., adjacent-hour) accumulation window over the period extending from the later of 19 Oct 2021 00:00 UTC or the forecast start time to the earlier of 30 Oct 2021 00:00 UTC or the forecast end time—that is, from *max*(19 Oct 2021 00:00 UTC, forecast start time) to *min*(30 Oct 2021 00:00 UTC, forecast end time). The largest value across all N-hour realizations was then retained at each grid point.

**Precipitation depth being an accumulated quantity, it was computed from the 0–6 h forecast interval rather than from the 0-h field, which is always zero.*

## III. Dynamical downscaling and application of the initial and boundary condition shifting method

In this study, we selected two ST targets: (i) the Willamette River watershed in Oregon (~11,500 mi$^2$), located approximately 6° north and 2° west of the region of heaviest precipitation associated with the observed AR cluster (Northern California Coast Ranges and Northern Sierra Nevada; see Fig. 3d–f); and (ii) the Nass River watershed in British Columbia (~8,000 mi$^2$), located approximately 16° north and 8° west of that region. Both watersheds therefore lie beyond the IBCS shift limits of ±1° in longitude and ±5° in latitude, which were suggested as acceptable bounds by Toride et al. (2019).

Among the 2,193 members of the IVL ensemble constructed in the previous section, several realizations directly impacted each target watershed—showing that IVL can transpose an AR to basins hundreds to over a thousand kilometers away, without any artificial modification of the regional model's IBCs, and can generate not just one, but several parallel, physically consistent and plausible realizations of the storm's evolution over the target. Among those realizations, the one producing the largest 24-h BA PD was selected for each watershed. A 24-h accumulation window was adopted because the local impact of the 24–25 October AR—the most intense in the observed cluster and the one responsible for most of its total precipitation—generally did not exceed about one day (as can be seen by comparing Fig. 3e to 3f, or 3h to 3i), as its precipitation field migrated southward during this period. However, an AR's duration, as well as its local rainfall persistence, can vary substantially from event to event (Ralph et al., 2013, 2019), so other accumulation durations could also have been pertinent. A 3-day window, for example, has commonly been used in previous PB PMP estimation studies (e.g., Ishida et al., 2015a;

Toride et al., 2019). The appropriate choice ultimately depends on the PMP duration one seeks to estimate.

Coincidentally, both of these realizations—those associated with the largest 24-h BA PDs (one per watershed)—originate from the same forecast cycle, initialized at 00:00 UTC on 17 October 2021. Their PW and IVT fields at 00:00 UTC on 28 October 2021 are shown in Fig. 4. Examination of the forecasted AR clusters reveals noticeable deviations from the observed sequence. Both realizations produced five ARs, compared to six observed between 19 and 30 October. Whereas the observed 24–25 October AR was preceded by a weaker AR on 23 October, the forecasted clusters exhibit only a single AR making landfall near 18:00 UTC on 23 October in both cases. What occurred afterwards differs substantially between the two forecasts.

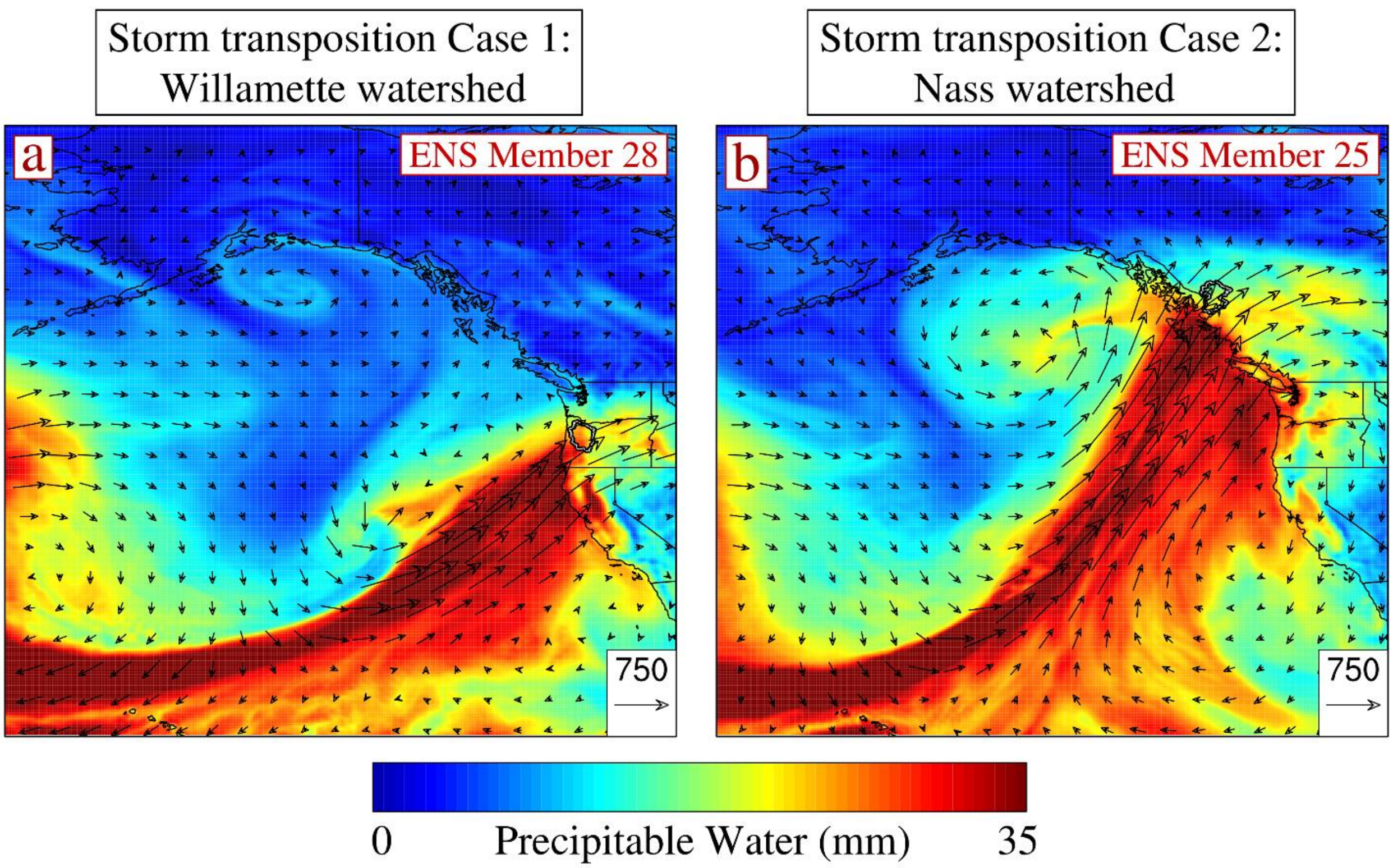

**Figure 4** – PW and IVT (kg m$^{-1}$ s$^{-1}$) fields at 28 Oct 2021 00:00 UTC from the two ECMWF forecast realizations that produced the largest 24-h BA PDs (among the 2193 realizations) in the two ST cases considered: (a) Willamette River watershed (member 28 of the forecast initialized at 17 Oct 2021 00:00 UTC), and (b) Nass River watershed (member 25 of the same forecast cycle).

In the Willamette case, the fourth and fifth forecasted ARs resembled the fifth and sixth observed ARs in that they consisted of two successive ARs originating from the same parent moisture plume. However, the forecasted ARs made landfall in northern California around 12:00 UTC on 26 October and in Oregon around 00:00 UTC on 28 October—the latter being the event that generated the maximum 24-h BA PD (Fig. 4a). By contrast, the fifth and sixth observed ARs were weaker and made landfall, respectively, in northern California around 00:00 UTC on 27 October and in Washington State around 12:00 UTC on 28 October. In the Nass case, the 23 October AR was followed by two distinct events: one making landfall in the Bay Area around 18:00 UTC on 25 October, and another, very intense AR striking western British Columbia around 00:00 UTC on 28 October (Fig. 4b), which generated the largest 24-h BA PD.

The fact that two perturbed members of a single forecast cycle can yield AR clusters with such different structures and timings, and capable of affecting such widely separated regions, is a striking illustration of the atmosphere's nonlinear behavior and of the potential of IVL for ST. Interestingly, at 00:00 UTC on 30 October—the end of the selected analysis period—both forecasts exhibit a sixth AR that is about to make landfall: in the Pacific Northwest for the Willamette case, and in western British Columbia for the Nass case. This indicates that the end date of the analysis period adopted in this study was somewhat premature, since, at least in these two realizations—and likely in many others among the 2,193—the AR cluster continues beyond 30 October 00 UTC, with an additional AR forming during the last two days of the month. The cutoff at 30 October 00:00 UTC was initially chosen because ERA5 only showed a weak AR reaching Alaska around 12:00 UTC that day, and it was not anticipated that internal variability could reorganize the large-scale environment to yield strong ARs impacting the target regions after 30 October 00:00 UTC. Extending the analysis window through the end of October would therefore have captured more ARs within the cluster, thereby providing additional plausible candidates for ST toward the Willamette and Nass watersheds.

To further enhance the 24-h BA PDs obtained from the two selected realizations, the IBCS method is subsequently applied. Before proceeding, however, the regional model's ability to reproduce the observed AR cluster's intense PD field must first be evaluated. As the 24–25 October AR contributed most of the cluster's total precipitation, the evaluation focused on the reconstruction of that event's PD field. In this study, DD was performed using the WRF model, version 4.6.1 (Skamarock et al., 2019). Three sets of nested computational domains were used (Fig. 5): one for the application of IBCS to each of the two ST cases described above, and one dedicated to reconstructing the PD field of the 24–25 October AR for model validation. The corresponding WRF model configurations—including physical parameterizations, grid dimensions and resolutions, and simulation start and end dates— are summarized in Table 1. We also note that the pressure-level, near-surface, and land-surface fields required for running WRF (see Section 2) were regridded from the original O640 reduced Gaussian grid to a 0.125° × 0.125° regular latitude–longitude grid for use with the WRF Preprocessing System (WPS).

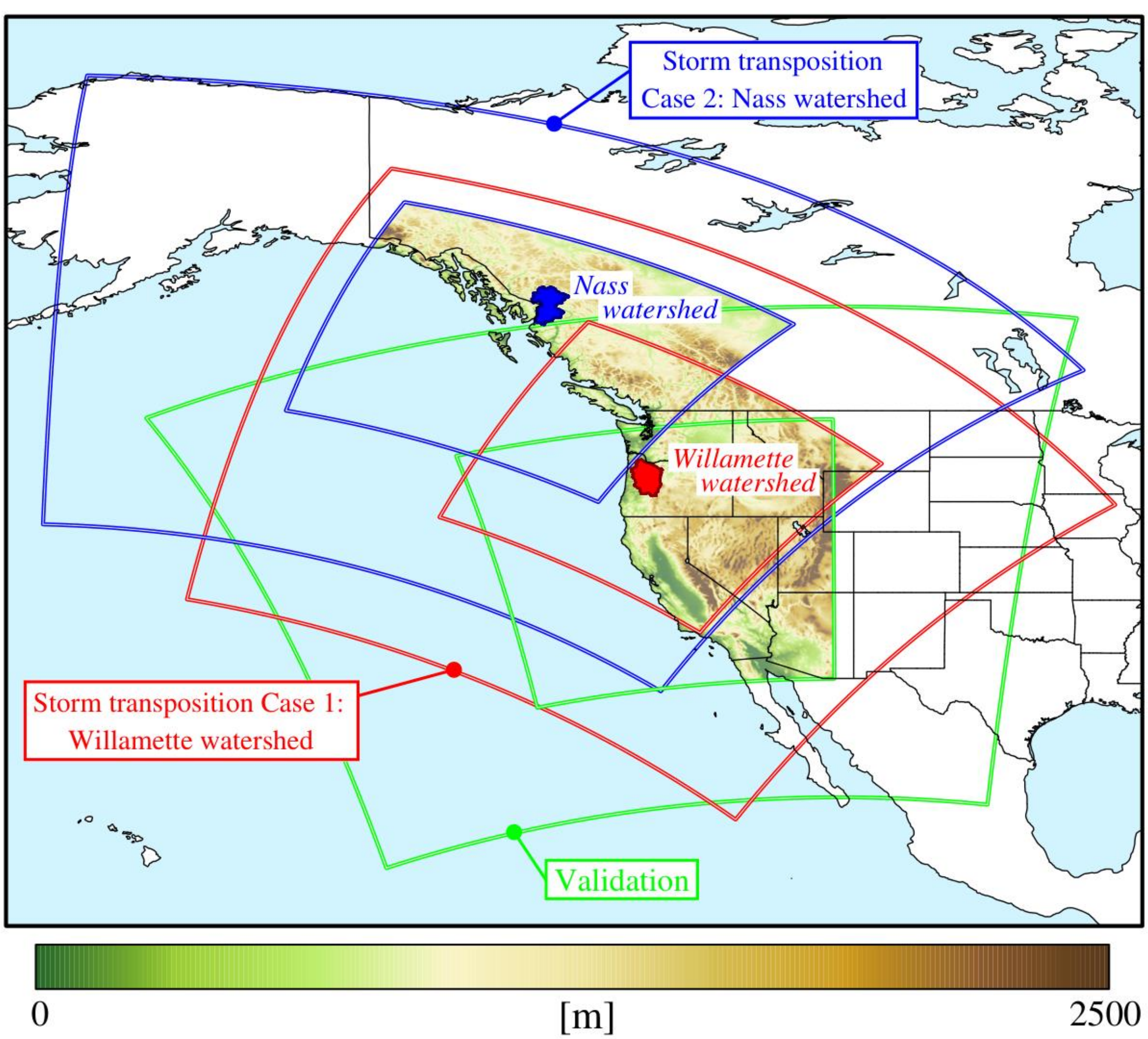

**Figure 5** – WRF model domain configuration. The green frames show the boundaries of the nested domains used for DD of the ECMWF ENS control forecast at 0-h lead time, to assess model performance in simulating the PD field of the 24-25 Oct 2021 AR. The red and blue frames show the nested domain boundaries for the two ST cases considered in this paper, in which the Willamette River watershed and the Nass River watershed, respectively, were selected as ST targets, with IBCs obtained from two perturbed members of the ECMWF

medium-range ensemble. The background shading in the inner domain shows terrain elevation (m) at 4-km resolution, obtained from the WRF model outputs.

**Table 1** – WRF model configuration for DD, including temporal and spatial resolution, grid dimensions, and physical parameterizations. All times are in Coordinated Universal Time (UTC).

| | **WRF model validation** | **ST Case 1: Willamette** | **ST Case 2: Nass** |
|---|---|---|---|
| **Simulation start time** | Outer: 2021-10-21 00:00<br>Inner: 2021-10-23 00:00 | Outer: 2021-10-23 06:00<br>Inner: 2021-10-25 06:00 | Outer: 2021-10-24 06:00<br>Inner: 2021-10-26 06:00 |
| **Simulation end time** | 2021-10-27 00:00 | 2021-10-29 18:00 | 2021-10-29 18:00 |
| **Map projection** | Lambert Conformal | | |
| **Grid dimensions (nx x ny)** | Outer: 360 x 305<br>Inner: 478 x 478 | Outer: 330 x 280<br>Inner: 433 x 403 | Outer: 330 x 280<br>Inner: 433 x 403 |
| **Grid spacing** | Outer: 12 km<br>Inner: 4 km | | |
| **Vertical levels** | 44 | | |
| **Model top pressure** | 50 hPa | | |
| **Time step** | Outer: 60 s<br>Inner: 20 s | | |
| **Initial & boundary conditions source** | Outer: ECMWF ENS control forecast at 0-h lead time<br>Inner: Outer domain | Outer: ECMWF ENS, member 28, initialized at 2021-10-17 00:00<br>Inner: Outer domain | Outer: ECMWF ENS, member 25, initialized at 2021-10-17 00:00<br>Inner: Outer domain |
| **Microphysics** | WRF Single-Moment 6-class graupel scheme (Hong & Lim, 2006) | | |
| **Cumulus** | Outer: Grell-3 (Grell and Dévényi, 2002)<br>Inner: None | | |
| **Planetary Boundary Layer** | Mellor-Yamada Nakanishi and Niino (MYNN) Level 2.5 (Nakanishi & Niino, 2006) | | |
| **Longwave Radiation** | Rapid Radiative Transfer Model (RRTM; Mlawer et al., 1997) | | |
| **Shortwave Radiation** | Dudhia (Dudhia, 1989) | | |
| **Land Surface** | Noah Land Surface Model (Mitchell et al., 2004) | | |
| **Surface Layer** | MYNN (Nakanishi & Niino, 2006) | | |

For the reconstruction with WRF of the PD field of the 24–25 October AR, we used the ENS control forecast at 0-h lead time for the IBCs. Although ERA5 could have been employed for model validation, we opted for consistency with the IBC data used in the ST cases—namely, perturbed members 28 (Willamette) and 25 (Nass) from the forecast cycle initialized at 00:00 UTC on 17 October 2021, as described previously. The resulting PDs for the period 24 October 00:00 UTC to 26 October 00:00 UTC were compared against Stage IV observations. Results are shown in Figs. 6 and 7, which display the spatial distribution of the PD fields and their empirical cumulative distribution functions (CDFs), respectively. These figures also include PDs calculated from the initial 0–6 h segment[8] of the ENS control forecast, so as to once again verify the accuracy of the ECMWF precipitation forecasts in the context of this AR event.

A comparison of Figs. 6a and 6c, as well as of the green and blue curves in Fig. 7, indicates that the ECMWF IFS notably underestimated the most intense precipitation associated with this AR, although the location and overall pattern of the PD field were well captured. This confirms that the intensity of the composite PD fields in Figs. 3a–c is likely underestimated (as discussed in Section 2), whereas their spatial structure can reasonably be considered reliable. Furthermore, Fig. 6b shows that DD led to substantial improvements in reproducing the intense precipitation from the AR, although the very most extreme PDs remain slightly underestimated, as indicated by the uppermost tail of the red and blue curves in Fig. 7.

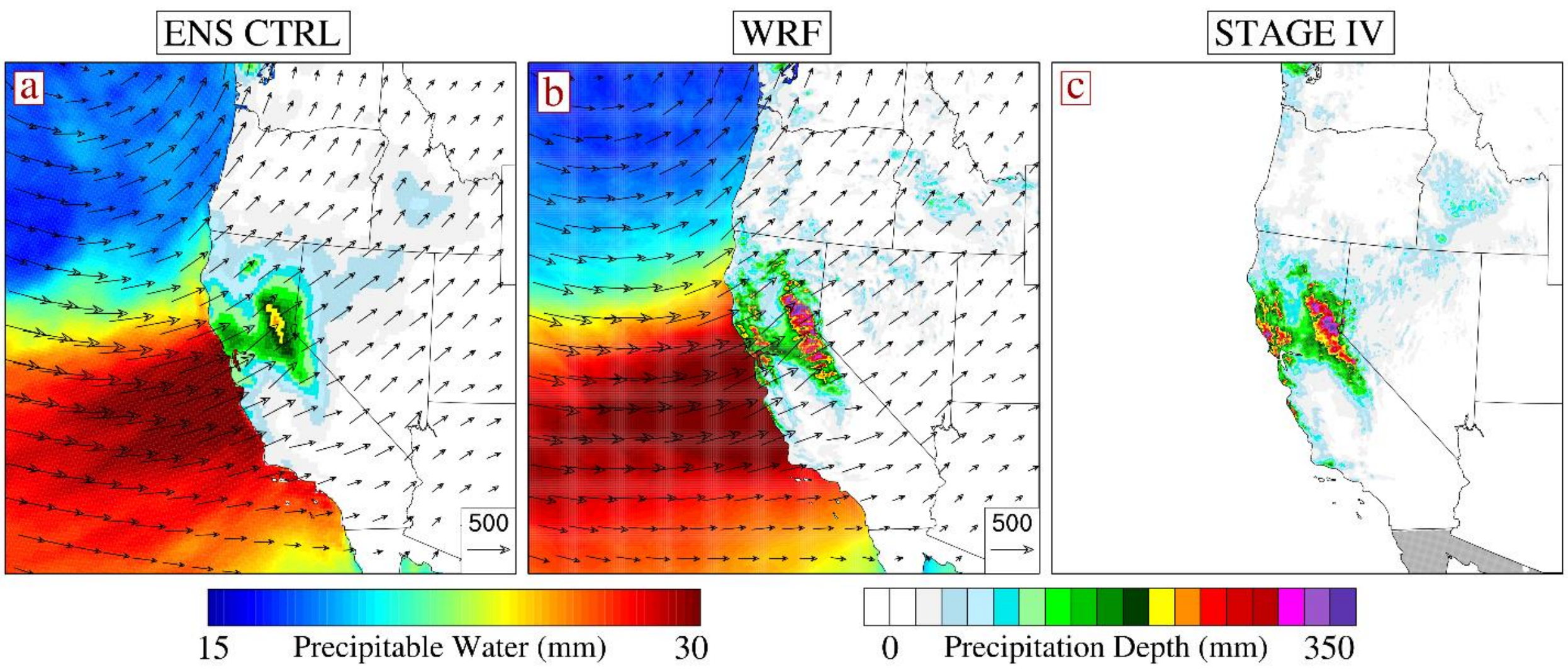

**Figure 6** – Overland PD (mm) fields accumulated from 24 Oct 2021 00:00 to 26 Oct 2021 00:00 UTC from: (a) the ECMWF ENS control forecast at 0-h lead time*, (b) the WRF downscaling of the ECMWF ENS control forecast at 0-h lead time (corresponding to the "WRF model validation" column in Table 1), and (c) Stage IV observations. Panels (a) and (b) also show PW (mm) over the ocean and IVT (kg $m^{-1}$ $s^{-1}$) averaged over the same period.

[8] Precipitation depth being an accumulated quantity, it was computed from the 0–6 h forecast interval rather than from the 0-h field, which is always zero.

In panel (c), the grey shaded region marks areas where Stage IV data are unavailable. We note that the region shown in this figure corresponds to the green inner domain in Fig. 5.

* *To be precise, since PD is an accumulated quantity, the 48-h PD was computed from the 0–6 h forecast interval rather than the 0-h field (which is always zero). Additionally, in panel (a), the PD, PW and IVT fields were first regridded from the O640 reduced Gaussian grid of the archived data to a 0.125° × 0.125° regular latitude–longitude grid for plotting purposes.*

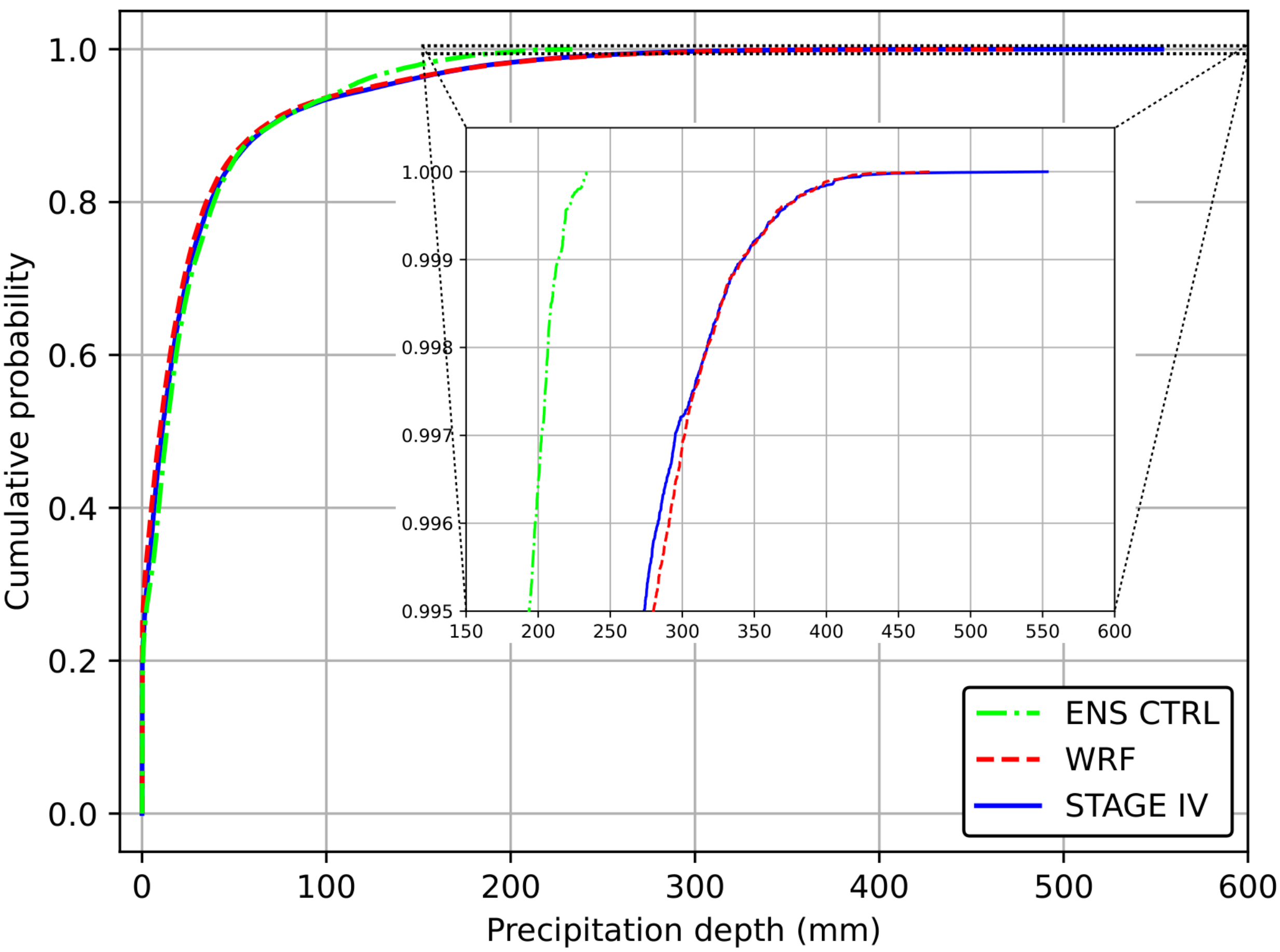

**Figure 7** – Empirical cumulative distribution functions (CDFs) of the 48-h PD fields from Fig. 6. To enable meaningful comparison, PD data from the ECMWF ENS control forecast (originally on an O640 reduced Gaussian grid) and from the WRF output (originally on a 3-km Lambert conformal grid) were conservatively regridded onto the Stage IV polar stereographic grid. Only grid points with available Stage IV observations were used in the CDF calculation. An inset highlights the upper tail of the distributions to emphasize differences in extreme precipitation.

DD with the WRF model was then performed for the two ECMWF ENS realizations that produced the largest 24-h BA PDs over the Willamette River and Nass River watersheds, and the IBCS method was applied to further enhance PD. As discussed in Section 2, the choice of the IBC shift limits used in IBCS is somewhat arbitrary and will likely continue to depend on the hydrometeorologist's or design engineer's judgment of what constitutes acceptable bounds. In this study, IBC shifts were limited to ±0.5° in both longitude and latitude—magnitudes that are considerably smaller than the distances separating the target watersheds from the areas most affected by precipitation from the observed AR cluster (see Fig. 3d–i).

IBCs were first subjected to shifts from a grid (circles in Fig. 8) spanning 0.5°W–0.5°E and 0.5°S–0.5°N, with a spacing of 0.1° in both directions. A subsequent refinement of the shifting amounts was performed around the three IBCS realizations producing the largest 24-h BA PDs (triangles in Fig. 8), followed by a second round of refinement (stars in Fig. 8). The maximum 24-h BA PD occurred either along the edge (Willamette: 0.25°E, 0.5°N) or in a corner (Nass: -0.5°E, 0.5°N) of the shift grid, suggesting that these PDs could potentially be further increased by exploring larger IBC shifts. However, we remained faithful to the initially defined IBC shift limits and did not explore larger-amplitude shifts. The application of IBCS enhanced the 24-h BA PDs from 104 mm to 119 mm over Willamette and from 90 mm to 98 mm over Nass. The corresponding PD fields are shown in Fig. 9.

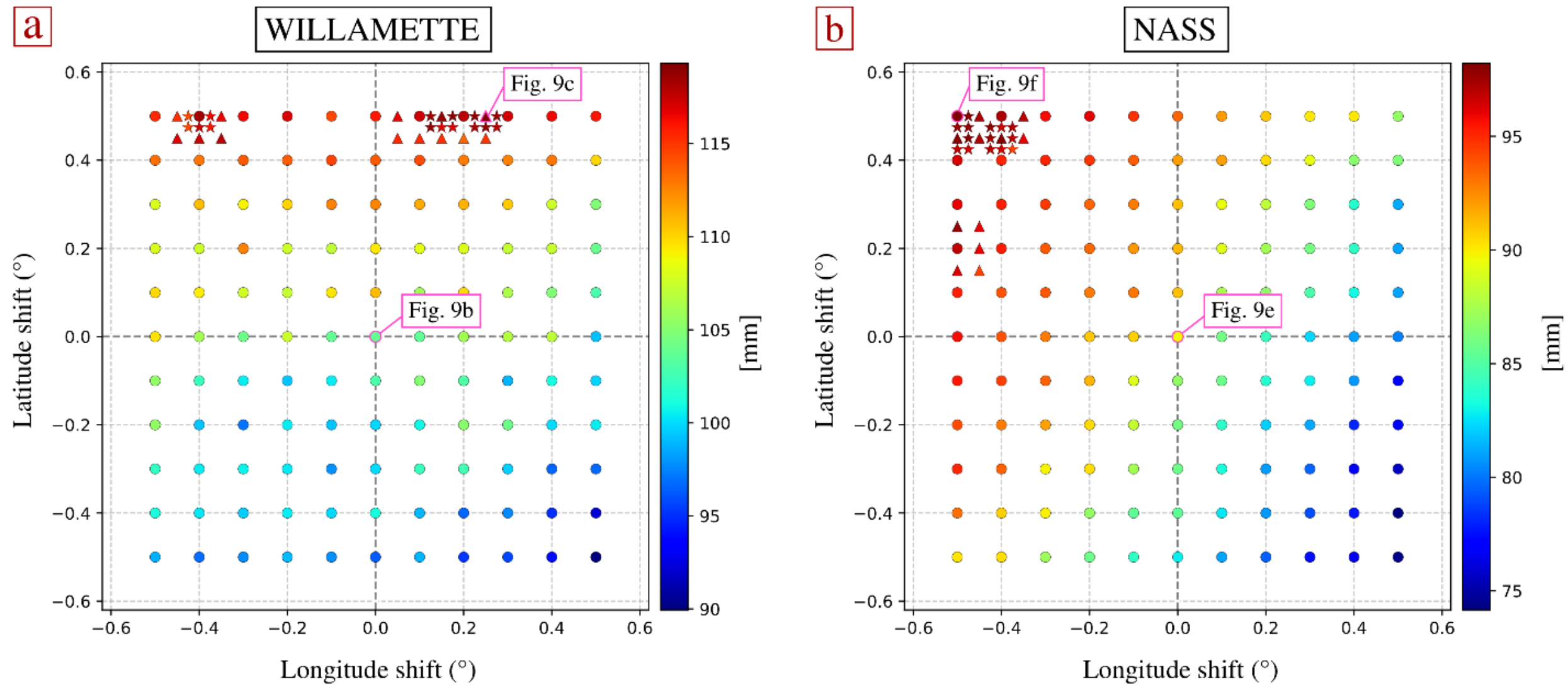

**Figure 8** – 24-h BA PD (mm) maxima as a function of IBC shifts (°) for (a) the Willamette River watershed case and (b) the Nass River watershed case, obtained by applying the IBCS method to the WRF downscaling of the two ECMWF forecast realizations that yielded the largest 24-h BA PDs. Each point of the plot represents one simulation, with x- and y-axes indicating the zonal and meridional shifts, respectively. Point color denotes the corresponding 24-h BA PD maximum, computed by sliding a 24-h accumulation window over the time series of hourly BA PDs between the inner-domain simulation start and end times (see Table 1). IBCS was first applied to a grid (shown by the circles) spanning 0.5°W–0.5°E and 0.5°S–0.5°N at 0.1° spacing in both directions. Triangles and stars indicate the first and second shift refinements, respectively.

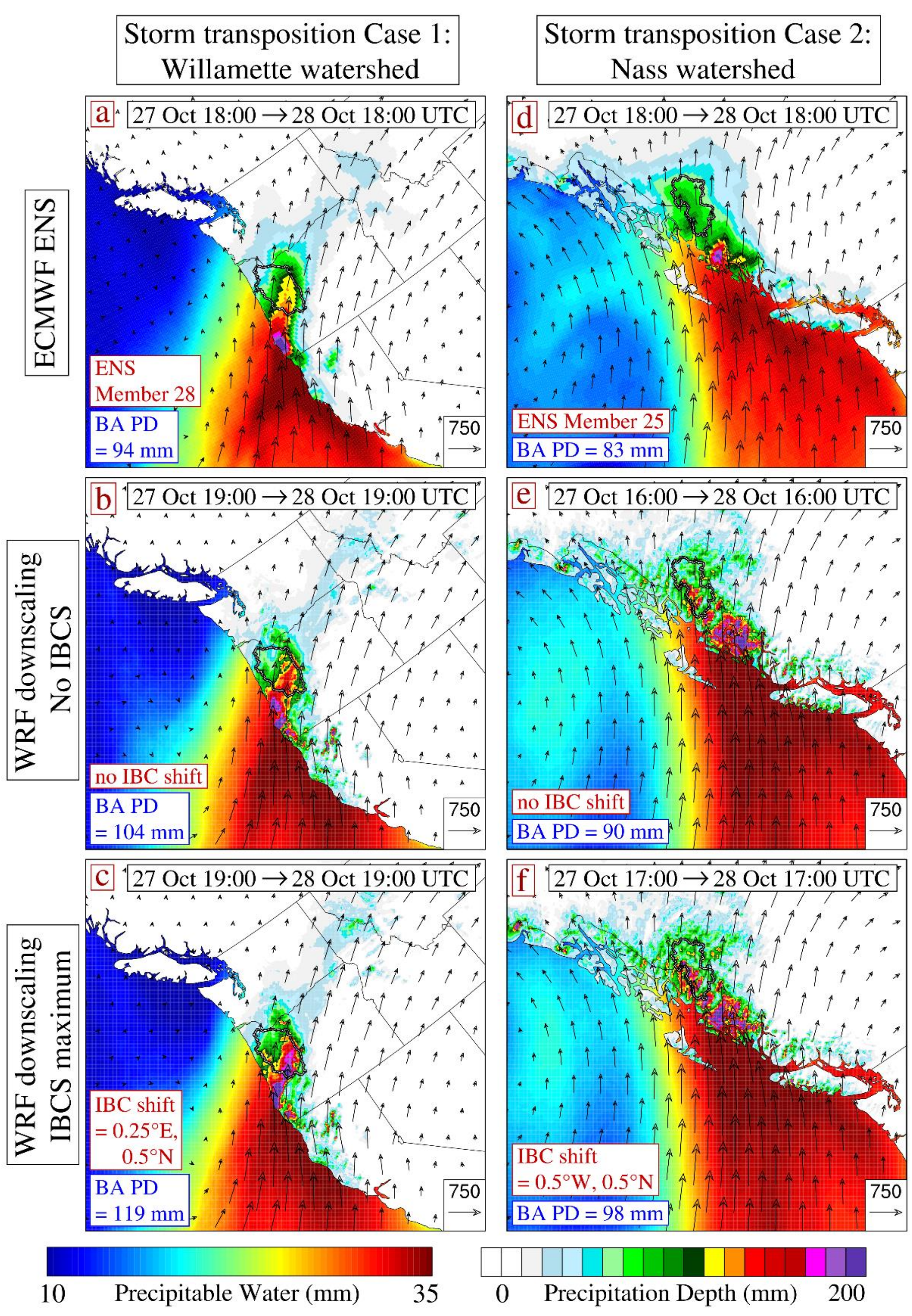

**Figure 9** – 24-h overland PD (mm) fields, and associated average PW (mm, over the ocean) and IVT (kg $m^{-1}$ $s^{-1}$) fields, for the two ECMWF ENS realizations that produced the largest 24-h BA PDs* (among the 2193 realizations) in the two ST cases considered: Willamette River watershed (maximum 24-h BA PD from member 28 of the forecast initialized at 17 Oct

2021 00:00 UTC) on the left, and Nass River watershed (maximum 24-h BA PD from member 25 of the same forecast cycle) on the right. The top row shows the ECMWF forecasts[**] for the 24-h period of maximum BA PD. The middle row shows the WRF-downscaled fields (setup in Table 1), using the 24-h window yielding the maximum BA PD from WRF outputs. The bottom row shows the IBCS realizations (highlighted in Fig. 8) that produced the largest BA PD. In each panel, the relevant 24-h period is indicated at the top, and the corresponding BA PD value is given at the bottom left. We note that the regions shown in this figure are the inner domains from Fig. 5: red for the Willamette watershed case and blue for the Nass watershed case.

[*] *The 24-h BA PD maximum of a given simulation was computed by sliding a 24-h accumulation window over the time series of hourly BA PDs between the inner-domain simulation start and end times (Table 1).*

[**] *In panels (a) and (d), the PD, PW and IVT fields were first regridded from the O640 reduced Gaussian grid of the archived data to a 0.125° × 0.125° regular latitude–longitude grid for plotting purposes.*

## IV. Summary and Conclusions

This article addresses storm transposition (ST)—the process by which a storm, whether historically observed or simulated, is displaced from its original location to a target region (such as a watershed) for the purpose of estimating the probable maximum precipitation (PMP), i.e., the greatest physically plausible precipitation depth over that area. The main technical challenge of this process lies in performing it in a physically and dynamically consistent manner, and, relatedly, in determining where it is legitimate to transpose the storm. Traditional hydrometeorological approaches rely on transposition factors that have weak scientific grounding and depend heavily on subjective judgment by PMP practitioners. The initial and boundary condition shifting (IBCS) method (Ishida et al., 2015a) represents a pioneering advance, marking a major technical shift in approaching ST by employing dynamical downscaling through kilometer-scale numerical weather prediction simulations. This approach ensures the conservation of mass, momentum, and energy within the model domains while fully accounting for the effects of topography. However, the simple shifting of atmospheric fields—which are intrinsically spatially nonstationary—can introduce physical inconsistencies in the regional simulation's initial and boundary conditions (IBCs). Moreover, the choice of acceptable IBC shift limits remains somewhat arbitrary, and the IBCS method does not provide a physically consistent criterion for defining the spatial extent of a storm's transposition region.

This article expands upon the work of Mure-Ravaud and Kavvas (2026), who proposed a new approach to ST. Although this framework still employs IBCS in its final stage to fine-tune the position of the storm's precipitation depth (PD) field over the target basin and enhance the basin-average (BA) PD, the "heavy lifting" in transporting the storm toward the target is performed by a process termed "internal variability leveraging" (IVL). In IVL, the intrinsic nonlinearity of the atmosphere—and its resulting sensitivity to perturbations in the IBCs—is exploited to generate an ensemble of plausible storm evolutions and trajectories, thereby enabling the storm to be transported to potentially distant locations in a manner that is fully physically consistent and does not rely on any artificial manipulation of the IBCs.

The mesoscale convective system investigated in MK26 had a relatively confined spatial extent, and its simulation with WRF revealed a pronounced sensitivity of the storm's structure and trajectory to the initialization and boundary configurations. Consequently, the lateral boundaries of the regional model imposed little constraint on the storm's ability to explore different regions of the inner domain. In this context, it was natural to use WRF itself to generate the IVL ensemble—that is, to leverage the model's own internal variability to produce ensemble members. This approach, however, was computationally demanding, as running over one hundred kilometer-scale simulations required several weeks of computation.

Limited-area models, however, leave little room for substantial ensemble spread in the case of more predictable, synoptic-scale systems such as atmospheric rivers (ARs), whose evolution is largely dictated by large-scale dynamics. For such systems, achieving a broader ensemble spread requires perturbations introduced further upstream in the modeling chain—at the level of the global model—following the approach of Ødemark et al. (2021).

Here, we construct a 2,193-member IVL ensemble of realizations of an AR cluster that affected the U.S. West Coast during 20–29 October 2021, directly from archived ECMWF medium-range forecasts. The advantage of this approach is that the IVL stage of the ST process requires no additional simulation, relying instead solely on the retrieval of existing forecast data from the archive. The limitation, however, lies in the reliability of PD data from the GCM. Here we found that ECMWF ENS PD forecasts for this AR cluster tend to be underestimated for the most extreme range of the PDs, although the location and structure of the PD fields are accurately represented. Dynamical downscaling with the WRF model at 4-km resolution, however, was found to bring substantial improvement in the simulation of the most intense PD associated with the AR cluster.

The initial intent of this study was to apply the IVL method only to the most intense AR within the observed cluster, on 24–25 October. However, we found that internal variability is such that, when considering earlier forecasts and different perturbed members, not only the landfall position is altered, but the very structure of the AR cluster may be affected. As a result, it becomes difficult to establish a one-to-one correspondence with the observed cluster, as the model begins to generate distinct, parallel AR sequences. The consequence of this—and one of the main findings of this study—is that the entire cluster must be treated as the operative unit for ST, since the large-scale dynamical and thermodynamical precursors to this cluster can unfold into a new configuration that may differ noticeably from the original.

Two ST targets were selected: the Willamette River watershed in northwestern Oregon and the Nass River watershed in western British Columbia. The two realizations that produced the largest 24-h BA PDs over these watersheds were then identified, and the IBCS method was applied to further enhance these PDs by fine-tuning the AR trajectory and the location of the resulting PD field. Application of the IVL–IBCS methodology ultimately yielded 24-h BA PDs of 119 mm over the Willamette and 98 mm over the Nass.

Future work will apply this methodology to other synoptic-scale storm systems to further evaluate its applicability and robustness.

## ACKNOWLEDGEMENTS

The author acknowledges the Hydrologic Research Laboratory (HRL) at the University of California, Davis for providing computational resources used to perform part of the simulations presented in this study.

The interpretations and conclusions expressed in this paper are solely those of the author and do not necessarily reflect the views of the Hydrologic Research Laboratory.

## CONFLICT OF INTEREST

The author has no relevant financial or non-financial interests to disclose.

## REFERENCES

Abbs, D. J. (1999). A numerical modeling study to investigate the assumptions used in the calculation of probable maximum precipitation. *Water Resources Research*, *35*(3), 785-796.

Ben Alaya, M. A., Zwiers, F., & Zhang, X. (2018). Probable maximum precipitation: Its estimation and uncertainty quantification using bivariate extreme value analysis. *Journal of Hydrometeorology*, *19*(4), 679-694.

Bowler, N. E., Arribas, A., Mylne, K. R., Robertson, K. B., & Beare, S. E. (2008). The MOGREPS short-range ensemble prediction system. *Quarterly Journal of the Royal Meteorological Society: A journal of the atmospheric sciences, applied meteorology and physical oceanography*, *134*(632), 703-722.

Buizza, R., Milleer, M., & Palmer, T. N. (1999). Stochastic representation of model uncertainties in the ECMWF ensemble prediction system. *Quarterly Journal of the Royal Meteorological Society*, *125*(560), 2887-2908.

Buizza, R., Bidlot, J. R., Wedi, N., Fuentes, M., Hamrud, M., Holt, G., & Vitart, F. (2007). The new ECMWF VAREPS (variable resolution ensemble prediction system). *Quarterly Journal of the Royal Meteorological Society: A journal of the atmospheric sciences, applied meteorology and physical oceanography*, *133*(624), 681-695.

Buizza, R., Leutbecher, M., & Isaksen, L. (2008). Potential use of an ensemble of analyses in the ECMWF Ensemble Prediction System. *Quarterly Journal of the Royal Meteorological Society: A journal of the atmospheric sciences, applied meteorology and physical oceanography*, *134*(637), 2051-2066.

Buizza, R. (2019). Ensemble generation: The TIGGE and S2S ensembles. In *Sub-seasonal to seasonal prediction* (pp. 261-303). Elsevier.

Charron, M., Pellerin, G., Spacek, L., Houtekamer, P. L., Gagnon, N., Mitchell, H. L., & Michelin, L. (2010). Toward random sampling of model error in the Canadian ensemble prediction system. *Monthly Weather Review*, *138*(5), 1877-1901.

Chen, J., Kavvas, M. L., Ishida, K., Trinh, T., Ohara, N., Anderson, M. L., & Chen, Z. R. (2016). Role of snowmelt in determining whether the maximum precipitation always results in the maximum flood. *Journal of Hydrologic Engineering*, *21*(10), 04016032.

Chen, L. C. (2005). *An investigation of the moisture maximization for the probable maximum precipitation*. The University of Iowa.

Chen, L. C., & Bradley, A. A. (2006). Adequacy of using surface humidity to estimate atmospheric moisture availability for probable maximum precipitation. *Water Resources Research*, *42*(9).

Chen, L. C., & Bradley, A. A. (2007). How does the record July 1996 Illinois rainstorm affect probable maximum precipitation estimates? *Journal of Hydrologic Engineering*, *12*(3), 327-335.

Coles, S. (2001). An Introduction to Statistical Modeling of Extreme Values. *London: Springer*.

Corrigan, P., Fenn, D. D., Kluck, D. R., & Vogel, J. L. (1999). Hydrometeorological Report No. 59: Probable Maximum Precipitation for California.

CW3E (Center for Western Weather and Water Extremes – Scripps Institution of Oceanography at UC San Diego). (2021). *CW3E event summary: 19–26 October 2021*. Retrieved October 8, 2025, from https://cw3e.ucsd.edu/cw3e-event-summary-19-26-october-2021

DeNeale, S. T., Kao, S.-C., & Watson, D. B. (2021). Considerations for Estimating Site-Specific Probable Maximum Precipitation at Nuclear Power Plants in the United States of America: Final Report (NUREG/KM-0015, ORNL/SPR-2021/1375). *United States Nuclear Regulatory Commission (NRC). Office of Nuclear Regulatory Research*. Available: https://www.nrc.gov/docs/ML2124/ML21245A418.pdf [Accessed: Feb. 13, 2025]

Dettinger, M. D. (2013). Atmospheric rivers as drought busters on the US West Coast. *Journal of Hydrometeorology*, *14*(6), 1721-1732.

Du, J. (2011). NCEP/EMC 4KM Gridded Data (GRIB) Stage IV Data. Version 1.0. UCAR/NCAR - Earth Observing Laboratory. https://doi.org/10.5065/D6PG1QDD

England, J. F. Jr., Sankovich, V. L., & Caldwell, R. J. (2020). Review of Probable Maximum Precipitation Procedures and Databases Used to Develop Hydrometeorological Reports (NUREG/CR-7131). *United States Nuclear Regulatory Commission (NRC). Office of Nuclear Regulatory Research*. Available: https://www.nrc.gov/docs/ML2004/ML20043E110.pdf [Accessed: Feb. 20, 2025]

Gimeno, L., Nieto, R., Vázquez, M., & Lavers, D. A. (2014). Atmospheric rivers: A mini-review. *Frontiers in Earth Science*, *2*, 2.

Guan, B., Molotch, N. P., Waliser, D. E., Fetzer, E. J., & Neiman, P. J. (2010). Extreme snowfall events linked to atmospheric rivers and surface air temperature via satellite measurements. *Geophysical Research Letters*, *37*(20).

Hansen, E. M., Schreiner, L. C., & Miller, J. F.  (1982). Hydrometeorological Report No. 52, Application of Probable Maximum Precipitation Estimates, United States East of the 105th Meridian. *National Weather Service.*

Hansen, E. M. (1987). Probable maximum precipitation for design floods in the United States. *Journal of Hydrology*, *96*(1-4), 267-278.

Hansen, E. M., Fenn, D. D., Schreiner, L. C., Stodt, R. W., & Miller, J. F. (1988). Hydrometeorological Report No. 55A, Probable Maximum Precipitation Estimates-United States between the Continental Divide and the 103rd Meridian. *National Weather Service*.

Hansen, E. M., Fenn, D. D., Corrigan, P., Vogel, J. L., Schreiner, L. C., & Stodt, R. W. (1994). Hydrometeorological Report No. 57, Probable Maximum Precipitation, Pacific Northwest States: Columbia River (including portions of Canada), Snake River and Pacific coastal drainages. *National Weather Service*.

Hersbach, H., Bell, B., Berrisford, P., Biavati, G., Horányi, A., Muñoz Sabater, J., Nicolas, J., Peubey, C., Radu, R., Rozum, I., Schepers, D., Simmons, A., Soci, C., Dee, D., Thépaut, J-N. (2023): ERA5 hourly data on pressure levels from 1940 to present. Copernicus Climate Change Service (C3S) Climate Data Store (CDS), DOI: 10.24381/cds.bd0915c6

Hersbach, H., Bell, B., Berrisford, P., Biavati, G., Horányi, A., Muñoz Sabater, J., Nicolas, J., Peubey, C., Radu, R., Rozum, I., Schepers, D., Simmons, A., Soci, C., Dee, D., Thépaut, J-N. (2023): ERA5 hourly data on single levels from 1940 to present. Copernicus Climate Change Service (C3S) Climate Data Store (CDS), DOI: 10.24381/cds.adbb2d47

Hershfield, D. M. (1961). Estimating the probable maximum precipitation. *Journal of the hydraulics Division*, *87*(5), 99-116.

Hershfield, D. M. (1965). Method for estimating probable maximum rainfall. *Journal-American Water Works Association*, *57*(8), 965-972.

Hiraga, Y., Iseri, Y., Warner, M. D., Frans, C. D., Duren, A. M., England, J. F., & Kavvas, M. L. (2021). Estimation of long-duration maximum precipitation during a winter season for large basins dominated by atmospheric rivers using a numerical weather model. *Journal of Hydrology*, *598*, 126224.

Hiraga, Y., Iseri, Y., Warner, M. D., Frans, C. D., Duren, A. M., England, J. F., & Kavvas, M. L. (2023). Comparison of Model-Based Precipitation Maximization Methods: Moisture Optimization Method, Storm Transposition Method, and Their Combination. *Journal of Hydrologic Engineering*, *28*(1), 04022033.

Hiraga, Y., & Meza, J. (2025a). Extreme precipitation modeling and Probable Maximum Precipitation (PMP) estimation in Chile. *Journal of Hydrology: Regional Studies*, *58*, 102274.

Hiraga, Y., Tahara, R., & Meza, J. (2025b). A methodology to estimate Probable Maximum Precipitation (PMP) under climate change using a numerical weather model. *Journal of Hydrology*, *652*, 132659.

Hiraga, Y., & Tahara, R. (2025c). Responses of Convective Heavy Rainfall to Atmospheric Moisture Amplification: Implications for Probable Maximum Precipitation Estimation. *Journal of Hydrometeorology*.

Hiraga, Y., Watanabe, S., Yamashita, T., & Takizawa, H. (2025b). Climate Change Effects on Probable Maximum Precipitation of Mesoscale Convective Systems: Model-based estimation and large ensemble-based frequency analysis. *Journal of Hydrology*, 133724.

Ho, F. P., & Riedel, J. T. (1980). Hydrometeorological Report No. 53: Seasonal Variation of 10-Square-Mile Probable Maximum Precipitation Estimates, United States East of the 105th Meridian. *U.S. Department of Commerce, National Oceanic and Atmospheric Administration, U.S. Nuclear Regulatory Commission. Silver Spring, MD*. Available: https://www.weather.gov/media/owp/hdsc_documents/PMP/HMR53.pdf [Accessed: Feb. 20, 2025].

Inverarity, G. W., Tennant, W. J., Anton, L., Bowler, N. E., Clayton, A. M., Jardak, M., ... & Wlasak, M. A. (2023). Met Office MOGREPS-G initialisation using an ensemble of hybrid four-dimensional ensemble variational (En-4DEnVar) data assimilations. Quarterly Journal of the Royal Meteorological Society, 149(753), 1138-1164.

Ishida, K., Kavvas, M. L., Jang, S., Chen, Z. Q., Ohara, N., & Anderson, M. L. (2015a). Physically based estimation of maximum precipitation over three watersheds in Northern California: Atmospheric boundary condition shifting. *Journal of Hydrologic Engineering*, *20*(4), 04014052.

Ishida, K., Kavvas, M. L., Jang, S., Chen, Z. Q., Ohara, N., & Anderson, M. L. (2015b). Physically based estimation of maximum precipitation over three watersheds in Northern California: Relative humidity maximization method. *Journal of Hydrologic Engineering*, *20*(10), 04015014.

Ishida, K., Kavvas, M. L., Chen, Z. R., Dib, A., Diaz, A. J., Anderson, M. L., & Trinh, T. (2018). Physically based maximum precipitation estimation under future climate change conditions. *Hydrological Processes*, *32*(20), 3188-3201.

Ishikawa, H., Oku, Y., Kim, S., Takemi, T., & Yoshino, J. (2013). Estimation of a possible maximum flood event in the Tone River basin, Japan caused by a tropical cyclone. *Hydrological Processes*, *27*(23), 3292-3300.

Kavvas, M. L., Mure-Ravaud, M., Dib, A., & Ishida, K. (2023). Numerical Modeling of Local Intense Precipitation Processes (NUREG/CR-7287). *United States Nuclear Regulatory Commission (NRC). Office of Nuclear Regulatory Research*. Available: https://www.nrc.gov/docs/ML2320/ML23207A195.pdf [Accessed: Feb. 13, 2025]

Klemeš, V. (2000). Tall tales about tails of hydrological distributions. I. *Journal of Hydrologic Engineering*, *5*(3), 227-231.

Klemeš, V. (2000). Tall tales about tails of hydrological distributions. II. *Journal of Hydrologic Engineering*, *5*(3), 232-239.

Koutsoyiannis, D. (1999). A probabilistic view of Hershfield's method for estimating probable maximum precipitation. *Water resources research*, *35*(4), 1313-1322.

Lavers, D. A., & Villarini, G. (2013). The nexus between atmospheric rivers and extreme precipitation across Europe. *Geophysical Research Letters*, *40*(12), 3259-3264.

Liang, J., & Yong, Y. (2022). Dynamics of probable maximum precipitation within coastal urban areas in a convection-permitting regional climate model. *Frontiers in Marine Science*, *8*, 747083.

Lin, Y., & Mitchell, K. E. (2005, January). 1.2 the NCEP stage II/IV hourly precipitation analyses: Development and applications. In *Proceedings of the 19th Conference Hydrology, American Meteorological Society, San Diego, CA, USA* (Vol. 10).

Lin, H., Gagnon, N., Beauregard, S., Muncaster, R., Markovic, M., Denis, B., & Charron, M. (2016). GEPS-based monthly prediction at the Canadian Meteorological Centre. *Monthly Weather Review*, *144*(12), 4867-4883.

Lorenz, E. N. (1963). Deterministic nonperiodic flow. *Journal of atmospheric sciences*, *20*(2), 130-141.

Lorenz, E. N. (1969). The predictability of a flow which possesses many scales of motion. *Tellus*, *21*(3), 289-307.

Madden, R. A. (1976). Estimates of the natural variability of time-averaged sea-level pressure. *Monthly Weather Review*, *104*(7), 942-952.

Molteni, F., Buizza, R., Palmer, T. N., & Petroliagis, T. (1996). The ECMWF ensemble prediction system: Methodology and validation. *Quarterly journal of the royal meteorological society*, *122*(529), 73-119.

Moran, P. A. P. (1957). The statistical treatment of flood flows. *Eos, Transactions American Geophysical Union*, *38*(4), 519-523.

Mure-Ravaud, M., Dib, A., Kavvas, M. L., Yegorova, E., & Kanney, J. (2019a). Physically based storm transposition of four Atlantic tropical cyclones. *Science of the Total Environment*, *666*, 252-273.

Mure-Ravaud, M., Kavvas, M. L., & Dib, A. (2019b). Impact of increased atmospheric moisture on the precipitation depth caused by Hurricane Ivan (2004) over a target area. *Science of the total environment*, *672*, 916-926.

Mure-Ravaud, M., & Kavvas, M. L. (2026). Physically-based transposition of a mesoscale convective system for estimating probable maximum precipitation. *Climate Dynamics*, *64*(3), 100.

NASA Earth Observatory. (2021). Extratropical cyclones drench West Coast. Retrieved October 8, 2025, from https://earthobservatory.nasa.gov/images/148999/extratropical-cyclones-drench-west-coast

National Academies of Sciences, Engineering, and Medicine (NASEM). 2024. *Modernizing Probable Maximum Precipitation Estimation*. Washington, DC: The National Academies Press. https://doi.org/10.17226/27460.

Nelson, B. R., Prat, O. P., Seo, D. J., & Habib, E. (2016). Assessment and implications of NCEP Stage IV quantitative precipitation estimates for product intercomparisons. *Weather and Forecasting*, *31*(2), 371-394.

Ødemark, K., Müller, M., & Tveito, O. E. (2021). Changing lateral boundary conditions for probable maximum precipitation studies: a physically consistent approach. *Journal of Hydrometeorology*, *22*(1), 113-123.

Ohara, N., Kavvas, M. L., Kure, S., Chen, Z. Q., Jang, S., & Tan, E. (2011). Physically based estimation of maximum precipitation over American River watershed, California. *Journal of Hydrologic Engineering*, *16*(4), 351-361.

Ohara, N., Kavvas, M. L., Anderson, M. L., Chen, Z. Q., & Ishida, K. (2017). Characterization of extreme storm events using a numerical model–based precipitation maximization procedure in the Feather, Yuba, and American River watersheds in California. *Journal of Hydrometeorology*, *18*(5), 1413-1423.

Oku, Y., Yoshino, J., Takemi, T., & Ishikawa, H. (2014). Assessment of heavy rainfall-induced disaster potential based on an ensemble simulation of Typhoon Talas (2011) with controlled track and intensity. *Natural Hazards and Earth System Sciences*, *14*(10), 2699-2709.

Paltan, H., Waliser, D., Lim, W. H., Guan, B., Yamazaki, D., Pant, R., & Dadson, S. (2017). Global floods and water availability driven by atmospheric rivers. *Geophysical Research Letters*, *44*(20), 10-387.

Powers, J. G., Klemp, J. B., Skamarock, W. C., Davis, C. A., Dudhia, J., Gill, D. O., ... & Duda, M. G. (2017). The weather research and forecasting model: Overview, system efforts, and future directions. *Bulletin of the American Meteorological Society*, *98*(8), 1717-1737.

Ralph, F. M., Neiman, P. J., & Wick, G. A. (2004). Satellite and CALJET aircraft observations of atmospheric rivers over the eastern North Pacific Ocean during the winter of 1997/98. *Monthly weather review*, *132*(7), 1721-1745.

Ralph, F. M., Neiman, P. J., & Rotunno, R. (2005). Dropsonde observations in low-level jets over the northeastern Pacific Ocean from CALJET-1998 and PACJET-2001: Mean vertical-profile and atmospheric-river characteristics. *Monthly weather review*, *133*(4), 889-910.

Ralph, F. M., Dettinger, M. D., Cairns, M. M., Galarneau, T. J., & Eylander, J. (2018). Defining “atmospheric river”: How the Glossary of Meteorology helped resolve a debate. *Bulletin of the American Meteorological Society*, *99*(4), 837-839.

Ralph, F. M., Neiman, P. J., Wick, G. A., Gutman, S. I., Dettinger, M. D., Cayan, D. R., & White, A. B. (2006). Flooding on California's Russian River: Role of atmospheric rivers. *Geophysical Research Letters*, *33*(13).

Ralph, F. M., Neiman, P. J., Kiladis, G. N., Weickmann, K., & Reynolds, D. W. (2011). A multiscale observational case study of a Pacific atmospheric river exhibiting tropical—Extratropical connections and a mesoscale frontal wave. *Monthly Weather Review*, *139*(4), 1169-1189.

Ralph, F. M., Coleman, T., Neiman, P. J., Zamora, R. J., & Dettinger, M. D. (2013). Observed impacts of duration and seasonality of atmospheric-river landfalls on soil moisture and runoff in coastal northern California. *Journal of Hydrometeorology*, *14*(2), 443-459.

Ralph, F. M., Rutz, J. J., Cordeira, J. M., Dettinger, M., Anderson, M., Reynolds, D., ... & Smallcomb, C. (2019). A scale to characterize the strength and impacts of atmospheric rivers. *Bulletin of the American Meteorological Society*, *100*(2), 269-289.

Ralph, F. M., Dettinger, M. D., Schick, L. J., & Anderson, M. L. (2020). Introduction to atmospheric rivers. In *Atmospheric rivers* (pp. 1-13). Cham: Springer International Publishing.

Rastogi, D., Kao, S. C., Ashfaq, M., Mei, R., Kabela, E. D., Gangrade, S., ... & Anantharaj, V. G. (2017). Effects of climate change on probable maximum precipitation: A sensitivity study over the Alabama-Coosa-Tallapoosa River Basin. *Journal of Geophysical Research: Atmospheres*, *122*(9), 4808-4828.

Reinhart, B. J., & Berg, R. (2022, February). *Tropical Cyclone Report: Tropical Storm Wanda (AL212021), 30 October–7 November 2021*. National Hurricane Center. https://www.nhc.noaa.gov/data/tcr/AL212021_Wanda.pdf

Sanders, F., & Gyakum, J. R. (1980). Synoptic-dynamic climatology of the "bomb". *Monthly Weather Review*, *108*(10), 1589-1606.

Schreiner, L. C., & Riedel, J. T. (1978). Probable Maximum Precipitation Estimates, United States – East of the 105th Meridian. *U.S. Department of Commerce, National Oceanic and Atmospheric Administration, U.S. Department of the Army Corps of Engineers*. Available: https://www.weather.gov/media/owp/hdsc_documents/PMP/HMR51.pdf [Accessed: Feb. 20, 2025]

Skamarock, W. C., & Klemp, J. B. (2008). A time-split nonhydrostatic atmospheric model for weather research and forecasting applications. *Journal of computational physics*, *227*(7), 3465-3485.

Skamarock, W. C., et al. (2019). A Description of the Advanced Research WRF Model Version 4. Mesoscale and Microscale Meteorology Laboratory - National Center for Atmospheric Research. NCAR Technical Notes. NCAR/TN-556+STR.

Sodemann, H., & Stohl, A. (2013). Moisture origin and meridional transport in atmospheric rivers and their association with multiple cyclones. *Monthly Weather Review*, *141*(8), 2850-2868.

Sun, J., Xue, M., Wilson, J. W., Zawadzki, I., Ballard, S. P., Onvlee-Hooimeyer, J., ... & Pinto, J. (2014). Use of NWP for nowcasting convective precipitation: Recent progress and challenges. *Bulletin of the American Meteorological Society*, *95*(3), 409-426.

Tarouilly, E., Cannon, F., & Lettenmaier, D. P. (2023). Improving confidence in model-based probable maximum precipitation: how important is model uncertainty in storm reconstruction and maximization?. *Journal of Hydrometeorology*, *24*(2), 257-267.

Tarouilly, E. G., Holman, K. D., & Lettenmaier, D. P. (2024). A Physically Constrained Model-Based Moisture Amplification Approach for Probable Maximum Precipitation (PMP) Estimation. *Journal of Hydrometeorology*, *25*(9), 1407-1420.

Thompson, P. D. (1957). Uncertainty of initial state as a factor in the predictability of large scale atmospheric flow patterns. *Tellus*, *9*(3), 275-295.

Toride, K., Iseri, Y., Warner, M. D., Frans, C. D., Duren, A. M., England, J. F., & Kavvas, M. L. (2019). Model-based probable maximum precipitation estimation: How to estimate the worst-case scenario induced by atmospheric rivers?. *Journal of Hydrometeorology*, *20*(12), 2383-2400.

Toth, Z., & Kalnay, E. (1997). Ensemble forecasting at NCEP and the breeding method. *Monthly Weather Review*, *125*(12), 3297-3319.

Trinh, T., Iseri, Y., Diaz, A. J., Snider, E. D., Anderson, M. L., & Levent Kavvas, M. (2022a). Maximization of historical storm events over seven watersheds in Central/Southern Sierra Nevada by means of atmospheric boundary condition shifting and relative humidity optimization methods. *Journal of Hydrologic Engineering*, *27*(3), 04021051.

Vitart, F., & Robertson, A. W. (2019). Introduction: Why sub-seasonal to seasonal prediction (S2S)?. In *Sub-seasonal to seasonal prediction* (pp. 3-15). Elsevier.

Wick, G. A., Neiman, P. J., Ralph, F. M., & Hamill, T. M. (2013). Evaluation of forecasts of the water vapor signature of atmospheric rivers in operational numerical weather prediction models. *Weather and Forecasting*, *28*(6), 1337-1352.

WMO (World Meteorological Organization). (1973). Manual for Estimation of Probable Maximum Precipitation. *First Edition. Geneva, Switzerland: World Meteorological Organization*

WMO. (1986). Manual for Estimation of Probable Maximum Precipitation. *Second Edition. Geneva, Switzerland: World Meteorological Organization*

WMO (2009). Manual on Estimation of Probable Maximum Precipitation (PMP). WMO-No. 1045

Yang, L., & Smith, J. (2018). Sensitivity of extreme rainfall to atmospheric moisture content in the arid/semiarid southwestern United States: Implications for probable maximum precipitation estimates. *Journal of Geophysical Research: Atmospheres*, *123*(3), 1638-1656.

Zhao, W., Smith, J. A., & Bradley, A. A. (1997). Numerical simulation of a heavy rainfall event during the PRE-STORM Experiment. *Water Resources Research*, *33*(4), 783-799.

Zhou, X., Zhu, Y., Hou, D., Luo, Y., Peng, J., & Wobus, R. (2017). Performance of the new NCEP Global Ensemble Forecast System in a parallel experiment. *Weather and Forecasting*, *32*(5), 1989-2004.

Zhu, Y., & Newell, R. E. (1994). Atmospheric rivers and bombs. *Geophysical Research Letters*, *21*(18), 1999-2002.

Zhu, Y., & Newell, R. E. (1998). A proposed algorithm for moisture fluxes from atmospheric rivers. *Monthly weather review*, *126*(3), 725-735.